\documentclass[aps,preprint,tightenlines,nofootinbib,superscriptaddress,byrevtex]{revtex4}
\usepackage{amsmath,amssymb,amsbsy}
\usepackage{graphicx}
\usepackage{enumerate}
\usepackage{comment}
\usepackage{color}
\definecolor{orange}{rgb}{1,0.5,0}

\def\str{{\mathrm{str}}}

\newcommand{\coarse}{$b\approx0.125$~fm }
\newcommand{\fine}{$b\approx0.09$~fm }

\begin{document}

\unitlength=1mm

% Greek Letters
\def\a{{\alpha}}
\def\b{{\beta}}
\def\d{{\delta}}
\def\D{{\Delta}}
\def\e{{\epsilon}}
\def\g{{\gamma}}
\def\G{{\Gamma}}
\def\k{{\kappa}}
\def\l{{\lambda}}
\def\L{{\Lambda}}
\def\m{{\mu}}
\def\n{{\nu}}
\def\w{{\omega}}
\def\O{{\Omega}}
\def\S{{\Sigma}}
\def\s{{\sigma}}
\def\t{{\tau}}
\def\th{{\theta}}
\def\x{{\xi}}

\def\ol#1{{\overline{#1}}}

%slash's
\def\Dslash{D\hskip-0.65em /}
\def\dslash{{\partial\hskip-0.5em /}}
\def\vslash{{\rlap \slash v}}
\def\qbar{{\overline q}}

% Jargon
\def\CPT{{$\chi$PT}}
\def\QCPT{{Q$\chi$PT}}
\def\PQCPT{{PQ$\chi$PT}}
\def\tr{\text{tr}}
\def\str{\text{str}}
\def\diag{\text{diag}}
\def\order{{\mathcal O}}
\def\vit{{\it v}}
\def\vD{\vit\cdot D}
\def\am{\alpha_M}
\def\bm{\beta_M}
\def\gm{\gamma_M}
\def\smb{\sigma_M}
\def\smt{\overline{\sigma}_M}
\def\tb{{\tilde b}}

\def\mc#1{{\mathcal #1}}

% Fields
\def\Bbar{\overline{B}}
\def\Tbar{\overline{T}}
\def\cBbar{\overline{\cal B}}
\def\cTbar{\overline{\cal T}}
\def\pq{(PQ)}

% PQ
\def\Dju{\Delta_{ju}}
\def\Drs{\Delta_{rs}}
\def\mju{m_{ju}}

\def\eqref#1{{(\ref{#1})}}

{\count255=\time\divide\count255 by 60 \xdef\hourmin{\number\count255}
  \multiply\count255 by-60\advance\count255 by\time
  \xdef\hourmin{\hourmin:\ifnum\count255<10 0\fi\the\count255}}

\preprint{NT-LBNL-13-001,
	UCB-NPAT-13-001,
	UNH-13-01} 

\vskip -1.5cm

\title{\bf The Scalar Strange Content of the Nucleon from Lattice QCD}

\author{P. M.~Junnarkar} 
	\affiliation{Department of Physics, University of New Hampshire, Durham, New Hampshire 03824-3568} 
\author{A.~Walker-Loud} 
	\affiliation{Nuclear Science Division, Lawrence Berkeley National Laboratory,\\ Berkeley, California 94720}
	\affiliation{Department of Physics, University of California,\\ Berkeley, California 94720}

%\date{\today\quad\hourmin}

\begin{abstract}
The scalar strange-quark matrix element of the nucleon is computed with lattice QCD.
A mixed-action scheme is used with domain-wall valence fermions computed on the staggered MILC sea-quark configurations.
The matrix element is determined by making use of the Feynman-Hellmann theorem which relates this strange matrix element to the change in the nucleon mass with respect to the strange-quark mass.
The final result of this calculation is $m_s \langle N | \bar{s} s |N \rangle = 49\pm10\pm15$~MeV and, correspondingly $f_s = m_s \langle N | \bar{s} s |N \rangle / m_N = 0.053\pm0.011\pm0.016$.

Given the lack of a quantitative comparison of this phenomenologically important quantity determined from various lattice QCD calculations, we take the opportunity to present such an average.
The resulting conservative determination is $f_s = 0.043\pm0.011$.
\end{abstract}

%\pacs{12.38.Gc}
\maketitle

%\pagebreak
%\tableofcontents
%\newpage

%%%%%%%%%%%%%%%%%%%%%%%%%%%%%%%%%%%%%%%%%%%%%%%%%%%
%
%	INTRODUCTION
%
%%%%%%%%%%%%%%%%%%%%%%%%%%%%%%%%%%%%%%%%%%%%%%%%%%%
\section{Introduction}
\noindent
Determining the strange content of the nucleon has been a long-standing interest of nuclear and  particle physicists.
The scalar strange content of the nucleon can be related to kaon-nucleon scattering and phenomenologically to the possible condensation of kaons in dense nuclear environments~\cite{Kaplan:1986yq,Nelson:1987dg}.
The strange content of the nucleon may also play an important role in the scattering of dark-matter particles off nuclei.
The general low-energy coupling of dark matter to nuclei has recently been worked out systematically using low-energy effective field theory~\cite{Fitzpatrick:2012ix,Fitzpatrick:2012ib}.
The spin-independent coupling is the simplest and has hence received the most attention historically.
The spin-independent elastic scattering of dark matter off a nucleon is proportional to the square of the scalar matrix elements $\langle N | m_q \bar{q} q |N \rangle$ for quarks of flavor $q$.~\cite{Bottino:1999ei,Bottino:2001dj,Kaplan:2000hh,Ellis:2008hf,Ellis:2009ai,Giedt:2009mr,Freytsis:2010ne,Hill:2011be,Cheung:2012qy}.
There are no direct experimental means of measuring these matrix elements.
The heavy quark $q=\{c,b,t\}$ matrix elements can be computed from perturbative QCD and are reasonably well known~\cite{Shifman:1978zn,Kryjevski:2003mh}.
The light quark $q=\{u,d\}$ matrix elements can be reasonably determined from $\pi N$ scattering~\cite{Koch:1982pu,Gasser:1990ce,Hoferichter:2012wf}.
The scalar strange-quark matrix element presents the most theoretical challenge to determine reliably and has contributed one of the largest uncertainties in dark-matter detection experiments~\cite{Bottino:1999ei,Bottino:2001dj,Ellis:2008hf} (cancellations between different contributions to potential dark-matter--matter cross sections lead to even larger uncertainty than previously appreciated~\cite{Hill:2011be}).
There have been estimates using baryon chiral perturbation theory and $SU(3)$ symmetry~\cite{Borasoy:1996bx} as well as constraints with earlier lattice calculations~\cite{Frink:2005ru}.
For these reasons, there has been a resurgent interest in determining $m_s \langle N | \bar{s} s | N \rangle$ using lattice QCD, beginning with the work in Refs.~\cite{Young:2009zb,Giedt:2009mr}.
It is more common in the context of dark-matter searches to normalize this quantity by the nucleon mass,
\begin{equation}\label{eq:fs}
f_s = \frac{m_s \langle N | \bar{s} s | N \rangle}{m_N}\, .
\end{equation}

There are two typical approaches used to determine this quantity from lattice QCD.
The scalar strange-quark matrix element can be directly computed
or one can take advantage of the Feynman-Hellmann theorem;
\begin{equation}\label{eq:FH}
m_s\langle N | \bar{s} s | N \rangle = m_s \frac{\partial m_N}{\partial m_s} \, .
\end{equation}
Most groups use the direct method~\cite{Takeda:2010cw,Babich:2010at,Bali:2011ks,Dinter:2012tt,Gong:2012nw,Oksuzian:2012rzb,Engelhardt:2012gd},
one group uses a hybrid approach which involves elements of both methods~\cite{Toussaint:2009pz,Freeman:2012ry},
and some groups use the Feynman-Hellmann method~\cite{Young:2009zb,Durr:2011mp,Horsley:2011wr,Semke:2012gs,Oksuzian:2012rzb,Shanahan:2012wh,Ren:2012aj,Jung:2013rz}.
For a recent review of the scalar strange content of the nucleon, see Ref.~\cite{Young:2013nn}.

The present work utilizes the Feynman-Hellmann theorem which has the following distinct advantages over the direct method: it is numerically less expensive and the ground state contributions to the two-point correlation functions can be significantly more reliably determined than plateaus in direct matrix element calculations with equal computing resources.

We begin by presenting details of our lattice calculation in Sec.~\ref{sec:latt_details}
and then present the determination of $m_s\langle N | \bar{s} s | N \rangle$ in Sec.~\ref{sec:sbars}.
We have found a quantitative comparison of various lattice QCD calculations of this quantity lacking in the literature.  Given its important phenomenological role, we were compelled to compile such a comparison, which we provide in Sec.~\ref{sec:results}, along with the results of the present work.
While lattice calculations of $f_s$ still need improvement, there is a welcoming consistency in the determination of this quantity from a wide variety of lattice calculations.

%%%%%%%%%%%%%%%%%%%%%%%%%%%%%%%%%%%%%%%%%%%%%%%%%%%
%
%	LATTICE RESULTS
%
%%%%%%%%%%%%%%%%%%%%%%%%%%%%%%%%%%%%%%%%%%%%%%%%%%%
\section{Details of the Lattice Calculation and Numerical Results\label{sec:latt_details}}
\noindent
The present work utilizes mixed-action lattice QCD calculations with domain-wall fermion~\cite{Kaplan:1992bt,Shamir:1992im,Shamir:1993zy,Shamir:1998ww,Furman:1994ky}
propagators computed on the $n_f = 2+1$ asqtad-improved~\cite{Orginos:1998ue,Orginos:1999cr}
rooted, staggered sea-quark configurations generated by the MILC Collaboration~\cite{Bernard:2001av,Bazavov:2009bb},
(with hypercubic-smeared~\cite{Hasenfratz:2001hp,DeGrand:2002vu,DeGrand:2003sf,Durr:2004as}
gauge links to improve the chiral symmetry properties of the domain-wall propagators),
a strategy initiated by the LHP Collaboration~\cite{Renner:2004ck,Edwards:2005kw,Edwards:2005ym,Hagler:2007xi,WalkerLoud:2008bp,Bratt:2010jn}.
A principal motivation for this choice is the good chiral symmetry properties of the domain-wall action, while utilizing the less numerically expensive lattice configurations of the staggered action.
It has been shown that the chiral symmetry properties of the valence domain-wall fermions highly suppresses sources of chiral symmetry breaking from the sea-quark action~\cite{Chen:2005ab,Chen:2006wf,Chen:2007ug,Chen:2009su}.
This has led to a number of important results, including
a determination of the kaon bag parameter $B_K$~\cite{Aubin:2009jh}; 
the charmed and static baryon spectrum~\cite{Lin:2009rx,Liu:2009jc}; 
charmed meson interactions with pions and kaons~\cite{Liu:2012zya};
hyperon axial charges~\cite{Lin:2007ap}; 
a number of results from the NPLQCD Collaboration including two-hadron scattering lengths~\cite{Beane:2005rj,Beane:2007xs,Beane:2006gj,Beane:2007uh,Beane:2006mx,Beane:2006gf,Torok:2009dg};
multi-meson interactions, condensates and the three-pion interaction~\cite{Beane:2007es,Detmold:2008fn,Detmold:2008yn};
as well as a number of others~\cite{Beane:2006pt,Beane:2006fk,Beane:2006kx,Detmold:2008bw}.
There have been a few other choices for mixed actions all utilizing overlap~\cite{Narayanan:1992wx,Narayanan:1994gw} valence-fermions on a variety of sea-quark configurations.
These include Wilson sea-fermions~\cite{Durr:2007ez},
twisted-mass sea-fermions~\cite{Cichy:2010ta,Cichy:2012vg},
domain-wall sea fermions~\cite{Li:2010pw,Lujan:2012wg},
and HISQ sea fermions~\cite{Basak:2012py}.
Mixed-action calculations are inherently unitarity violating with partially quenched effects only vanishing in the continuum limit.
It is therefore imperative to compare numerical results with the scaling violations predicted from the mixed-action effective field theory~\cite{Bar:2002nr,Bar:2003mh,Bar:2005tu,Golterman:2005xa,Tiburzi:2005is,Chen:2005ab,Prelovsek:2005rf,Aubin:2006hg,Chen:2006wf,Jiang:2007sn,Orginos:2007tw,Chen:2007ug,Aubin:2008wk,Chen:2009su}.
This has been undertaken to an exploratory extent with baryons~\cite{WalkerLoud:2008bp,Jenkins:2009wv}, but the only systematic studies have been with the $a_0$ correlator~\cite{Prelovsek:2005rf,Aubin:2008wk}, which is highly contaminated by the unitarity violating effects and a recent determination of low-energy constants in the two-flavor chiral Lagrangian for pions~\cite{Beane:2011zm}.
Despite the limited study of discretization effects, there are reasons to believe they are small for many quantities~\cite{Chen:2005ab,Chen:2006wf,Chen:2007ug,WalkerLoud:2008bp,WalkerLoud:2008pj}.

%%%%%%%%%%%%%%%%%%%%%%%%%%%%%%%%%%%%%%%%%%%%%%%%%%%
%
%	LATTICE PARAMETERS
%
%%%%%%%%%%%%%%%%%%%%%%%%%%%%%%%%%%%%%%%%%%%%%%%%%%%
\subsection{Parameters of the lattice QCD calculation}
\noindent
The present calculation utilizes the Feynman-Hellmann theorem to determine the scalar strange-quark matrix element in the nucleon, Eq.~\eqref{eq:FH},
limiting the work to a small set of available ensembles.
Details of the various ensembles and parameters are collected in Table~\ref{tab:latt_params}.
%%%%%%%%%%%%%%%%%%%%%%%%%%%%%%%%%%%%%%%%%%%%%%%%%%%
\begin{table}[t]
\caption{\label{tab:latt_params}
Parameters used in the present work.  For some of the calculations, the time direction was chopped at $t=32$ with Dirichlet boundary conditions (denoted by volumes with $\times32$).  For the MILC configurations, the notation m010m030 (and similar) means the input quark mass values are $bm_l=0.010$ and $bm_s = 0.030$ for the light and strange sea quarks respectively.
}
\begin{ruledtabular}
\begin{tabular}{cccccccccc}
%\hline
$\beta$& $m_{sea}$& V& $M_5$& $L_5$
& $bm_l^{dwf}$& $bm_l^{res}$& $bm_s^{dwf}$& $bm_s^{res}$
& $N_{src}\times N_{cfg}$\\
%&&&&&&$\times 10^3$\\
\hline
\multicolumn{10}{c}{$b \approx 0.125~\texttt{fm}$ ensembles}\\
6.75& m010m030& $20^3\times64$& $1.7$& $16$& 0.0138& 0.001564(03)& 0.081& 0.000892(2)& $53\times 328$\\
6.76& m010m050& $20^3\times64$& $1.7$& $16$& 0.0138& 0.001566(11)& 0.081& 0.000913(2)& $\phantom{0}4\times 656$\\
6.76& m010m050& $20^3\times32$& $1.7$& $16$& 0.0138& 0.001552(27)& 0.081& 0.000913(2)& $24\times 769$\\
6.79& m030m030& $20^3\times64$& $1.7$& $16$& 0.0478& 0.001052(04)& 0.081& 0.000809(4)& 
$30\times367$\\
6.81& m030m050& $20^3\times32$& $1.7$& $16$& 0.0478& 0.001013(06)& 0.081& 0.000862(7)& $24\times 564$\\
\hline
\multicolumn{10}{c}{$b \approx 0.09~\texttt{fm}$ ensembles}\\
%$\beta$& $m_{sea}$& V& $M_5$& $L_5$
%& $bm_l^{dwf}$& $bm_l^{res}$& $bm_s^{dwf}$& $bm_s^{res}$
%& $N_{src}\times N_{cfg}$\\
%\hline
7.08& m0031m0186& $40^3\times96$& $1.5$& $12$& 0.0035& 0.000431(3)& 0.0423& 0.000236(2)& $1\times 356$ \\
7.08& m0031m031& $40^3\times96$& $1.5$& $12$& 0.0035& 0.000428(3)& 0.0423& 0.000233(2)& $1\times 422$\\
\end{tabular}
\end{ruledtabular}
\end{table}
%%%%%%%%%%%%%%%%%%%%%%%%%%%%%%%%%%%%%%%%%%%%%%%%%%%
There are two sets of ensembles at the \coarse lattice spacing with fixed light-quark mass and strange-quark masses that straddle the physical strange-quark mass.  These are denoted by the sets $m_{sea} = \{\textrm{m010m030, m010m050}\}$ and $m_{sea} = \{\textrm{m030m030, m030m050}\}$, respectively.%
%FOOTNOTE
\footnote{The notation $m_{sea} =\textrm{m010m030}$ means the light quark has an input light quark mass value in lattice units of $bm_l = 0.010$ and the strange-quark input mass value is $bm_s = 0.030$.} 
On the \fine ensembles there are two sets, with fixed light quark mass and strange-quark masses straddling the physical strange-quark mass.  In this work, preliminary results are presented only for one of these sets with $m_{sea} = \{\textrm{m0031m0186, m0031m031}\}$.
The values of the domain-wall quark masses, the fifth-dimensional extent $L_5$, and the domain-wall mass $M_5$ were taken from the NPLQCD production runs~\cite{Beane:2011zm}.

%%%%%%%%%%%%%%%%%%%%%%%%%%%%%%%%%%%%%%%%%%%%%%%%%%%
%
%	 LATTICE RESULTS
%
%%%%%%%%%%%%%%%%%%%%%%%%%%%%%%%%%%%%%%%%%%%%%%%%%%%
\subsection{Results of the lattice calculation}
\noindent
The light- and strange-quark propagators were computed with a Gaussian-smeared source~\cite{Frommer:1995ik,Pochinsky:1997} and both smeared (SS) and point (PS) sinks.
Correlation functions were then constructed with the quantum numbers of the pion and proton.
The pion masses were determined with a fully correlated simultaneous fit to the SS and PS correlation functions,  with a single cosh used for both correlators,
\begin{align}
C_{(XS)}(t) \sim A_{(XS)}\  e^{-m_\pi T/2}\  \cosh ( m_\pi (t - T/2))\, ,
\end{align}
where $X=S,P$.  In all cases, the configurations are binned until the statistical uncertainty of the extracted masses stopped changing appreciably.
To determine the fitting systematic, the length of the time extent used in the fit and the starting time were varied over a wide range, with a minimum plateau length of $\sim0.5$~fm.
For each fit, the $Q$ value is used as a weight, where 
\begin{equation}\label{eq:Q}
	Q \equiv \int_{\chi^2_{min}}^{\infty} d \chi^2\ \mc{P}(\chi^2,d)\, ,
\end{equation}
with the probability distribution function for $\chi^2$ with $d$ degrees of freedom,
\begin{equation}
	\mc{P}(\chi^2,d) = \frac{1}{2^{d/2} \G(d/2)} (\chi^2)^{d/2 - 1} e^{-\chi^2 / 2}\, .
\end{equation}
The central value is determined from the weighted sum,
\begin{equation}
	\bar{m} = \frac{\sum_i m_i Q_i}{\sum_j Q_j}\, .
\end{equation}
In many cases, the systematic is approximately Gaussian, and so the 16\% and 84\% quantiles are used to determine the systematic uncertainties.

The choice to use the $Q$ values as weights is simply motivated.  $Q$ ranges from $[0,1]$ with a value of $1$ indicating the fit function and resulting parameters perfectly describe the correlation function over the range of fit.
It also allows one to compare fits with different model functions (e.g. single and double state fits).  While not the only choice for determining a fitting systematic, it is a convenient and useful choice.

\begin{figure}
\includegraphics[width=0.8\textwidth]{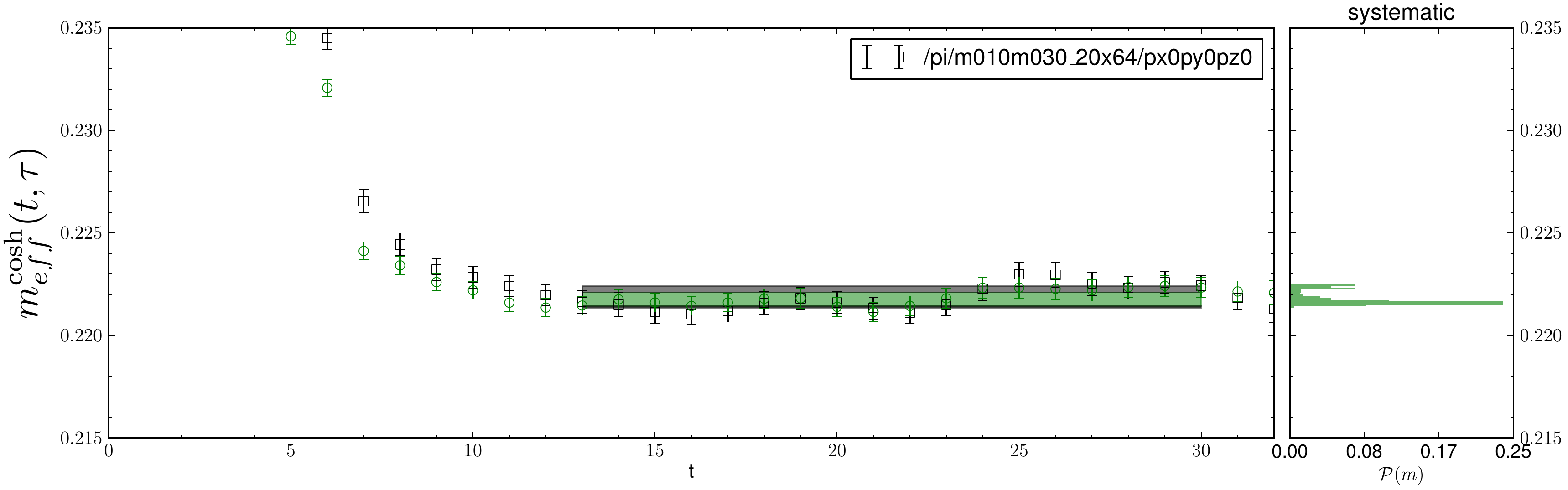}
\includegraphics[width=0.8\textwidth]{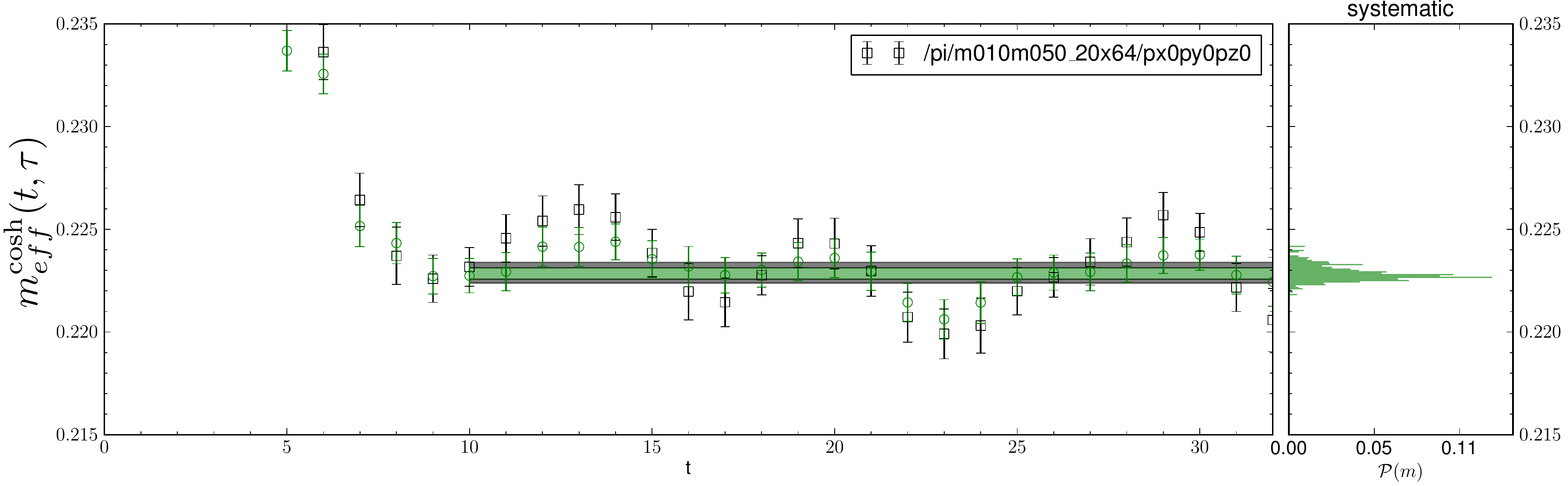}
\includegraphics[width=0.8\textwidth]{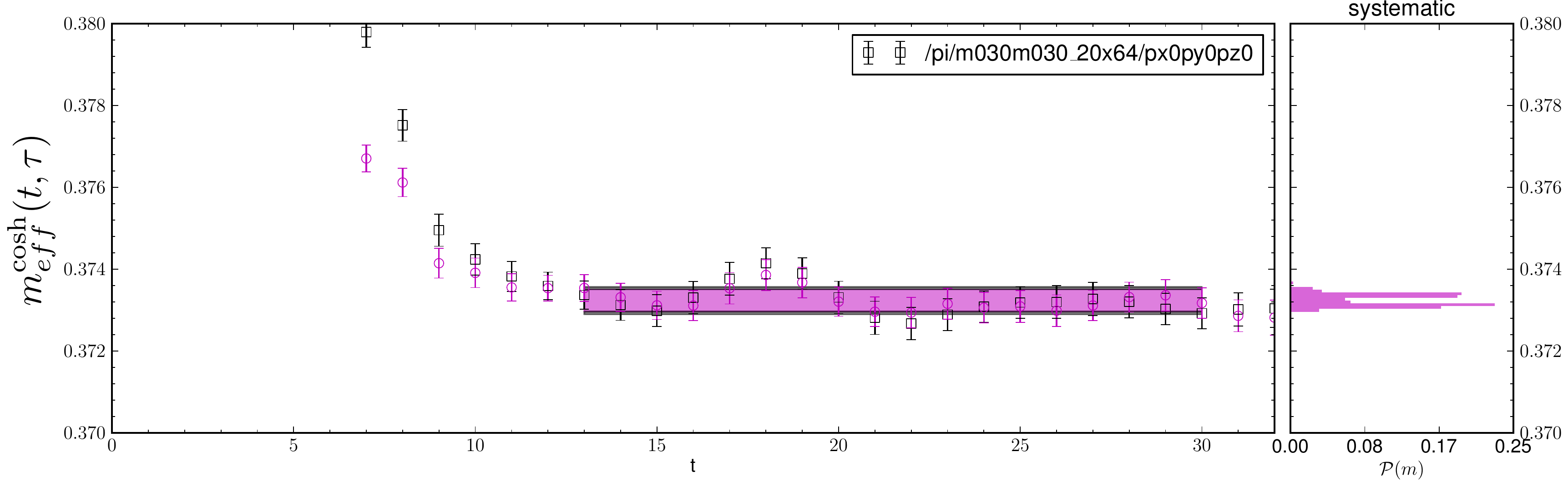}
\includegraphics[width=0.8\textwidth]{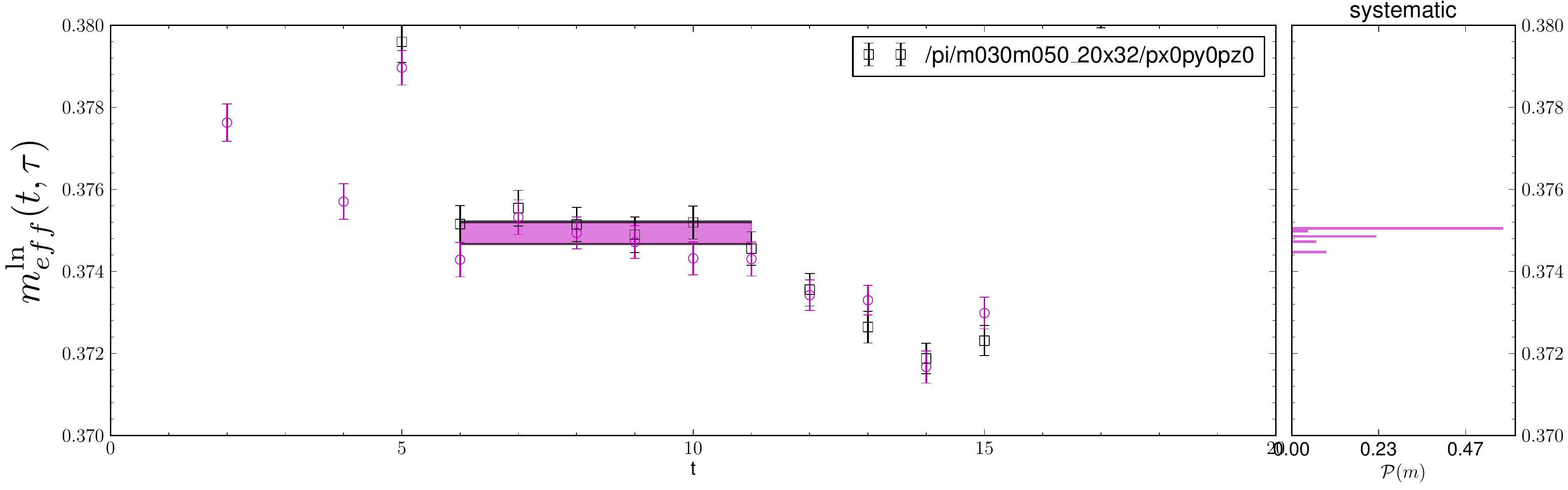}
\caption{\label{fig:mpi_coarse}Pion mass effective mass plots on the \coarse ensembles.}
\end{figure}

\begin{figure}
\includegraphics[width=0.8\textwidth]{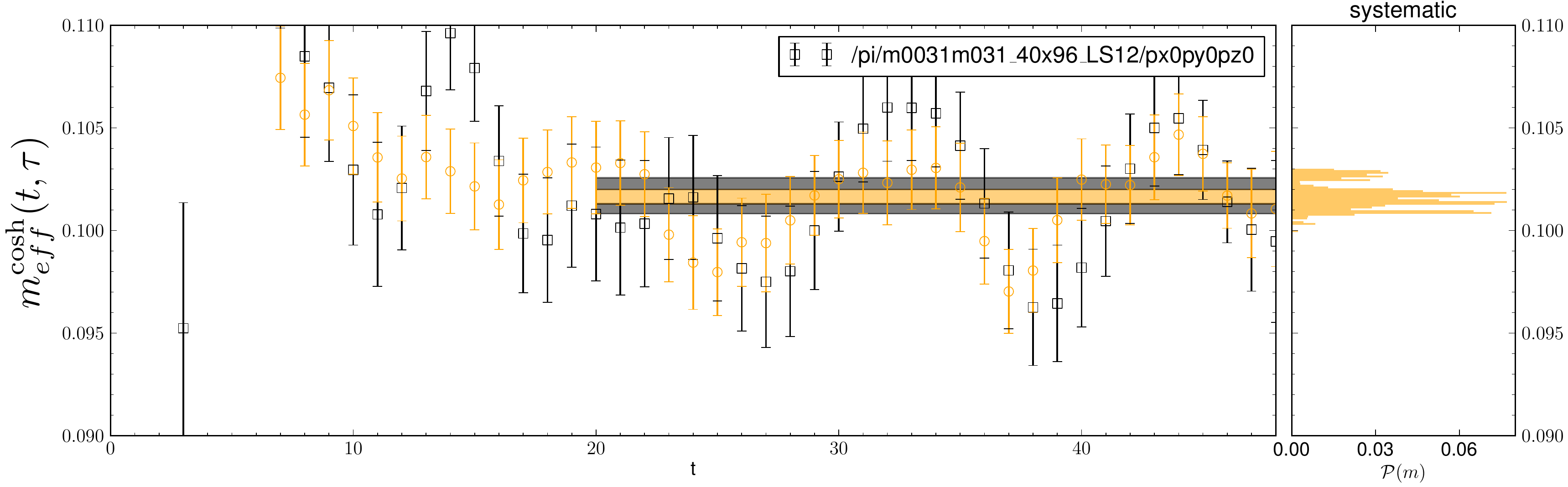}
\includegraphics[width=0.8\textwidth]{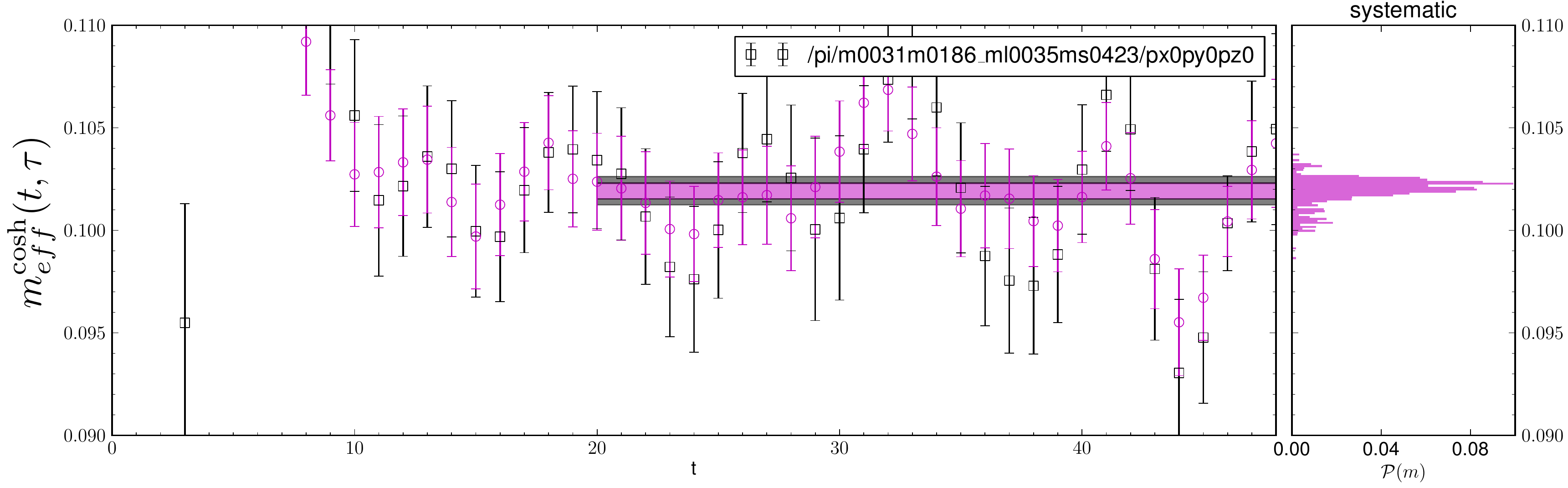}
\caption{\label{fig:mpi_fine}Pion mass effective mass plots on the \fine ensembles.}
\end{figure}

The results of these fits are plotted over a representative window in time along with cosh-style effective masses,
\begin{equation}
	m^{\cosh}_{eff}(t,\t) = \frac{1}{\t} \cosh^{-1} \left( 
		\frac{C(t+\t) + C(t-\t)}{C(t)} \right)
\end{equation}
in Figs.~\ref{fig:mpi_coarse} and \ref{fig:mpi_fine}.  The (black) squares are from the PS correlation functions while the (colored) open circles are from the SS correlation functions.
%%%%%%%%%%%%%%%%%%%%%%%%%%%%%%%%%%%%%%%%%%%%%%%%%%%
%%%%%%%%%%%%%%%%%%%%%%%%%%%%%%%%%%%%%%%%%%%%%%%%%%%
%%%%%%%%%%%%%%%%%%%%%%%%%%%%%%%%%%%%%%%%%%%%%%%%%%%
The right-side bar in each plot displays the mass probability distribution function determined from 
\begin{equation}
	\mc{P}_i(m) = \frac{Q_i}{\sum_j Q_j}\, .
\end{equation}
In all plots, the inner (colored) band represents the statistical uncertainty in the fit while the outer band represents the statistical and fitting systematic added in quadrature.
In the case of the $m_{sea} = \textrm{m030m050}$ ensemble, only results with Dirichlet boundary conditions in time are available.  For this case, the correlation functions are fit simultaneously with a single exponential,
\begin{align}\label{eq:corr_exp}
C_{(XS)}(t) \sim A_{(XS)}\  \exp ( - m_\pi t)\, ,
\end{align}
and compared with the standard effective mass,
\begin{equation}
	m^{\ln}_{eff}(t,\t) = \frac{1}{\t} \ln \left( \frac{C(t)}{C(t+\t)} \right)\, .
\end{equation}
Clearly, in this case, the ability to explore the fitting systematic is more limited.  For reasons discussed in Ref.~\cite{Beane:2011zm}, the dip in the effective mass is not believed to represent a lower ground state energy, but rather contaminations from the Dirichlet boundary condition.
The results are collected in Table~\ref{tab:mass_results}.
%%%%%%%%%%%%%%%%%%%%%%%%%%%%%%%%%%%%%%%%%%
%%  PIONS and NUCLEON masses
%%%%%%%%%%%%%%%%%%%%%%%%%%%%%%%%%%%%%%%%%%
\begin{table}
\caption{\label{tab:mass_results}Computed pion and nucleon masses on the various ensembles.  Additionally, the value of $r_1/b$ used to convert to physical units is provided, obtained from Refs.~\cite{Bazavov:2009bb,Bernard:r1b}.}
\begin{ruledtabular}
\begin{tabular}{cccccc}
$\beta$& $m_{sea}$& V& $bm_\pi$& $bm_N$& $\frac{r_1}{b}(bm_l^\textrm{phy},bm_s^\textrm{phy},\beta)$ \\
\hline
6.75& m010m030& $20^3\times64\times16$& $0.22178(33)({}^{54}_{28})$& $0.7177(18)({}^{19}_{26})$& 2.711(4) \\ 
6.76& m010m050& $20^3\times64\times16$& $0.22285(28)({}^{46}_{37})$& -- & 2.739(3) \\ 
6.76& m010m050& $20^3\times32\times16$& -- & $0.7311(19)({}^{36}_{26})$& 2.739(3) \\
6.79& m030m030& $20^3\times64\times16$& $0.37323(27)(20)$& $0.8653(17)({}^{27}_{33})$& 2.821(7) \\
6.81& m030m050& $20^3\times32\times16$& $0.37493(26)({}^{24}_{11})$& $0.8740(18)({}^{36}_{32})$& 2.877(4)\\
%\hline
7.06& m0031m0186& $40^3\times96\times12$& $0.10192(38)({}^{59}_{55})$& $0.4621(64)({}^{99}_{85})$& 3.687(4)\\
7.08& m0031m031& $40^3\times96\times12$& $0.10165(35)({}^{84}_{76})$& $0.4603(48)({}^{79}_{74})$& 3.755(4)
\end{tabular}
\end{ruledtabular}
\end{table}
%%%%%%%%%%%%%%%%%%%%%%%%%%%%%%%%%%%%%%%%%%

The proton masses are trickier to determine as the signal-to-noise ratio decays exponentially in time~\cite{Lepage:1989hd},
\begin{equation}
	\lim_{t\rightarrow \infty} \frac{S(t)}{N(t)} = A\, e^{-(m_N - \frac{3}{2}m_\pi)t}\, .
\end{equation}
The mass determined in a given fit from $t_i$ to $t_f$ is then susceptible to larger fitting systematics.
It is not uncommon for the effective mass plateau to shift by order one standard deviation and form a new plateau, either higher or lower at times when the statistical fluctuations grow appreciably.
It is therefore important to develop a systematic analysis algorithm that both takes advantage of the precise statistical fluctuations at early times while allowing for the possibility that the late-time fluctuations represent the true ground state.
In Ref.~\cite{Beane:2009kya}, it was demonstrated that correlation functions determined with $\mc{O}(10^5)$ reasonably statistically independent sources on $\mc{O}(10^4)$ Monte Carlo trajectories, a variety of analysis methods could be used all producing consistent results.
With fewer \textit{measurements}, not all methods work as well.
One technique which works better than others is the Matrix-Prony method~\cite{Prony} (similar to the variational method which has gained popularity lately), as described in Refs.~\cite{Beane:2009kya,Beane:2009gs}.
The general idea is to find linear combinations of correlation functions which isolate various eigenstates and allow for a determination of the masses starting from earlier Euclidean times.

The Matrix-Prony method is well suited to matrices of correlation functions that are neither square nor positive-definite, as is often the case in lattice QCD calculations.
One begins with the \textit{ansatz} that the (vector) of correlation functions can be described with a transfer matrix,
\begin{equation}\label{eq:MP_ansatz}
	y(t+\t) = \hat{T}(\t) y(t)\, ,
\end{equation}
where in our case $y(t)$ is composed of just two correlation functions,
\begin{equation}
	y(t) = \begin{pmatrix} C_{PS}(t)\\ C_{SS}(t) \end{pmatrix}\, .
\end{equation}
It is useful to factorize the transfer operator $\hat{T}(\t) = M^{-1}(\t) V$ and multiply on the right by the transpose vector to form the matrix equation,
\begin{equation}\label{eq:MP_2}
	M(\t) y(t+\t)y^T(t) = V y(t) y^T(t)\, .
\end{equation}
To be useful, Eq.~\eqref{eq:MP_ansatz} must be satisfied over a range of time,
\begin{equation}\label{eq:MP_3}
	M(\t) \sum_{t=t_0}^{t_0+\D t} y(t+\t)y^T(t) = V \sum_{t=t_0}^{t_0+\D t} y(t) y^T(t)\, .
\end{equation}
A solution to Eq.~\eqref{eq:MP_3} is given by
\begin{align}
&M(\t) = \left( \sum_{t=t_0}^{t_0+\D t} y(t+\t) y^T(t) \right)^{-1}\, ,&
&V = \left( \sum_{t=t_0}^{t_0+\D t} y(t) y^T(t) \right)^{-1}\, .&
\end{align}
In order to guarantee the inverse can be found, enough times must be summed over to ensure the corresponding matrices are of full rank.
One then solves the eigenvalue equation for the principal correlators,
\begin{align}
&\hat{T}(\t) q_n = (\l_n)^\t q_n\, ,&
&\textrm{with } \l_n = e^{-E_n}\, .&
\end{align}

A point that differentiates the Matrix-Prony method from other variational methods is the sum over time slices in Eq.~\eqref{eq:MP_3}.
Most variational methods pick a reference time at which to perform the diagonalization of the correlation functions, whereas with Matrix-Prony, one must sum over a number of time slices greater than or equal to the number of correlation functions.
Moreover, one can increase confidence in the subsequent analysis by maximizing $\D t$ in Eq.~\eqref{eq:MP_3}.
The original ansatz \eqref{eq:MP_ansatz} is satisfied if over the range of time, $t_0$ to $t_0+\D t$, the resulting principal correlation functions are well described by a single exponential.

%%%%%%%%%%%%%%%%%%%%%%%%%%%%%%%%%%%%%%%%%%%%%%%%%%%
\begin{figure}
\includegraphics[width=0.8\textwidth]{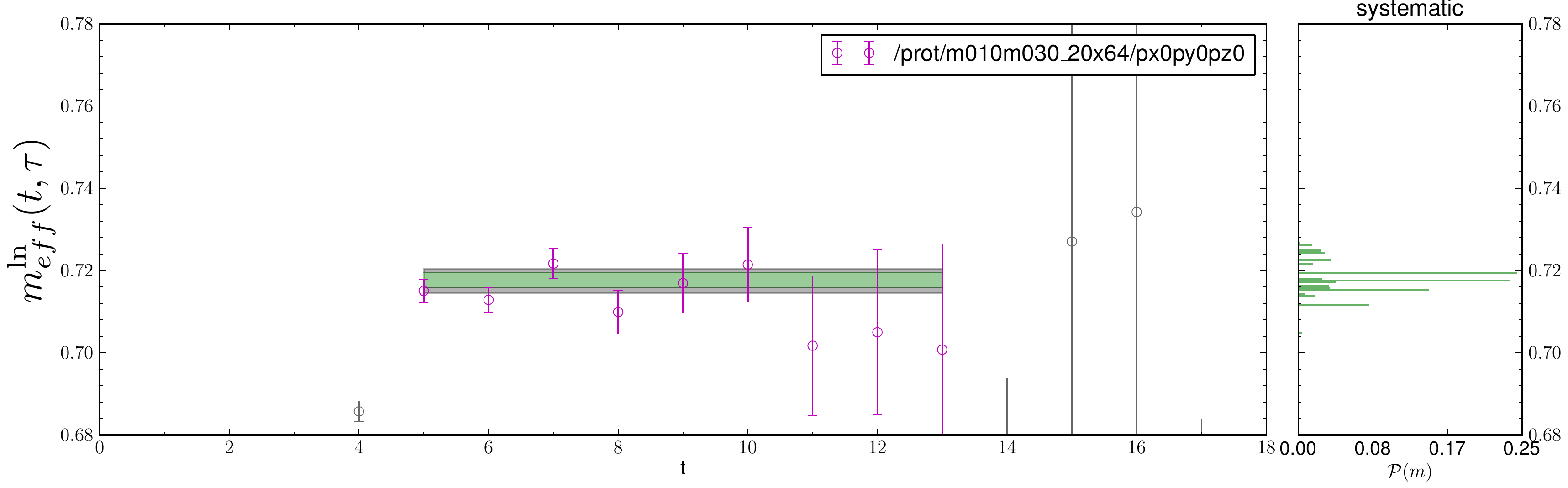}
\includegraphics[width=0.8\textwidth]{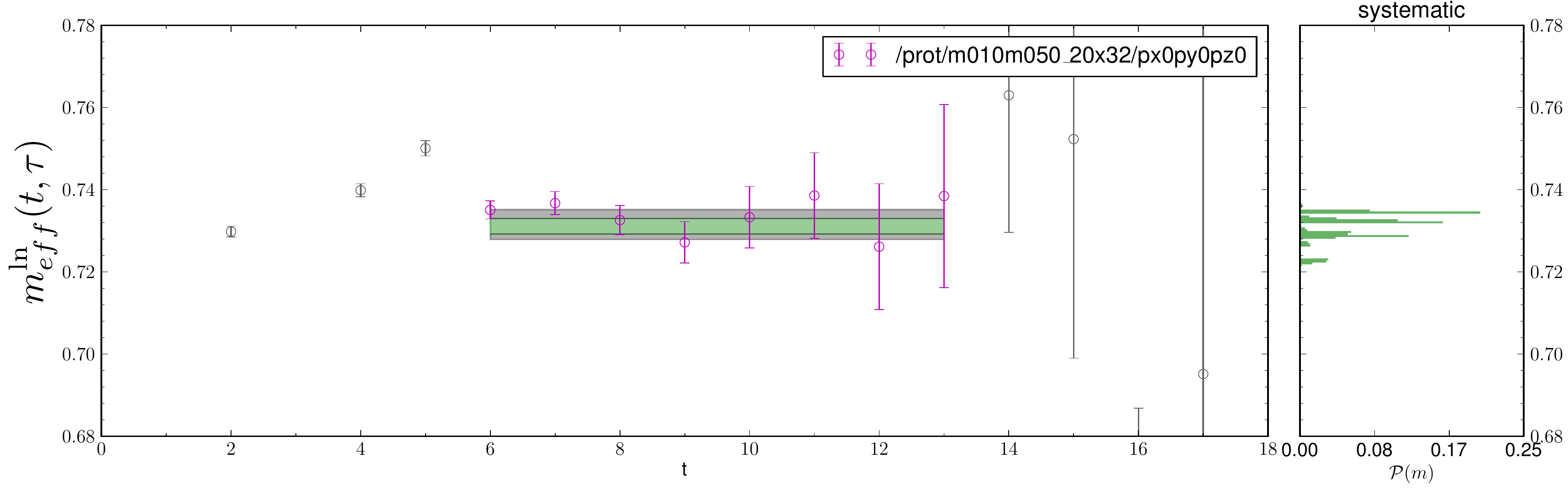}
\includegraphics[width=0.8\textwidth]{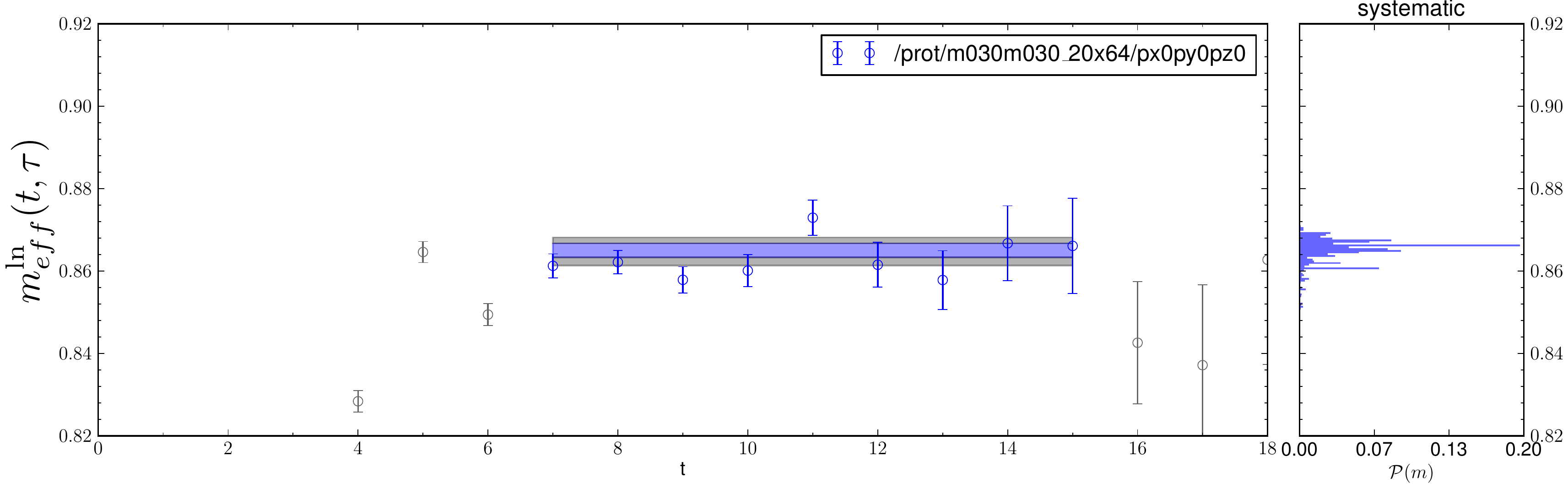}
\includegraphics[width=0.8\textwidth]{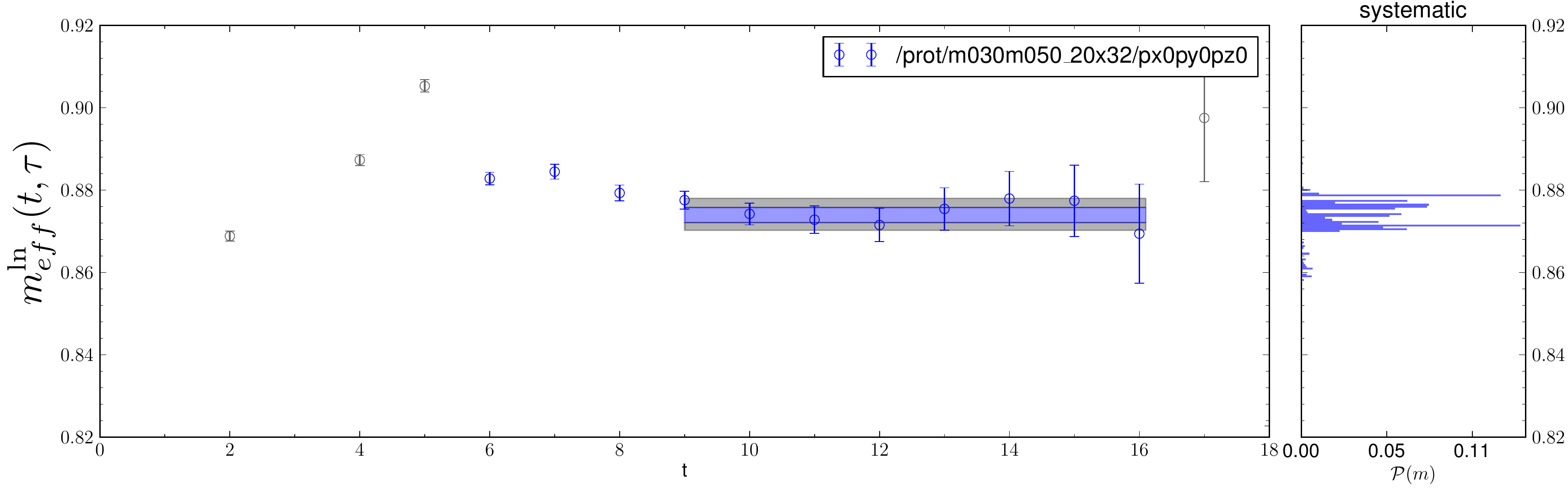}
\caption{\label{fig:mp_coarse}Proton mass and representative effective mass plots on the \coarse ensembles.}
\end{figure}
%%%%%%%%%%%%%%%%%%%%%%%%%%%%%%%%%%%%%%%%%%%%%%%%%%%
\begin{figure}
\includegraphics[width=0.8\textwidth]{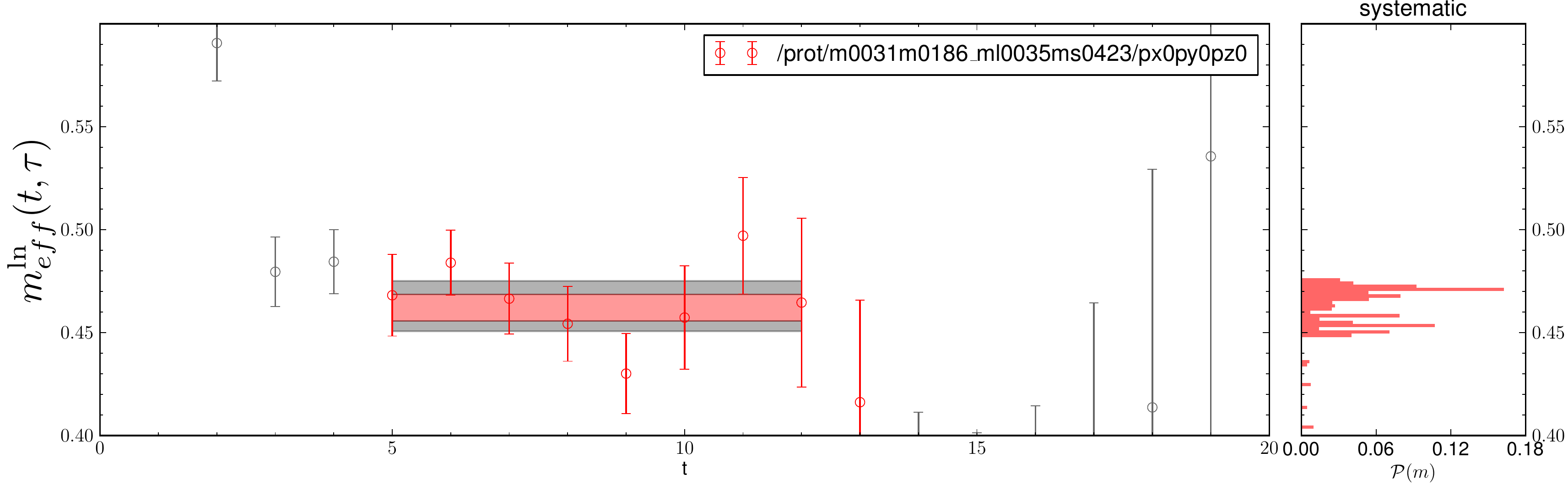}
\includegraphics[width=0.8\textwidth]{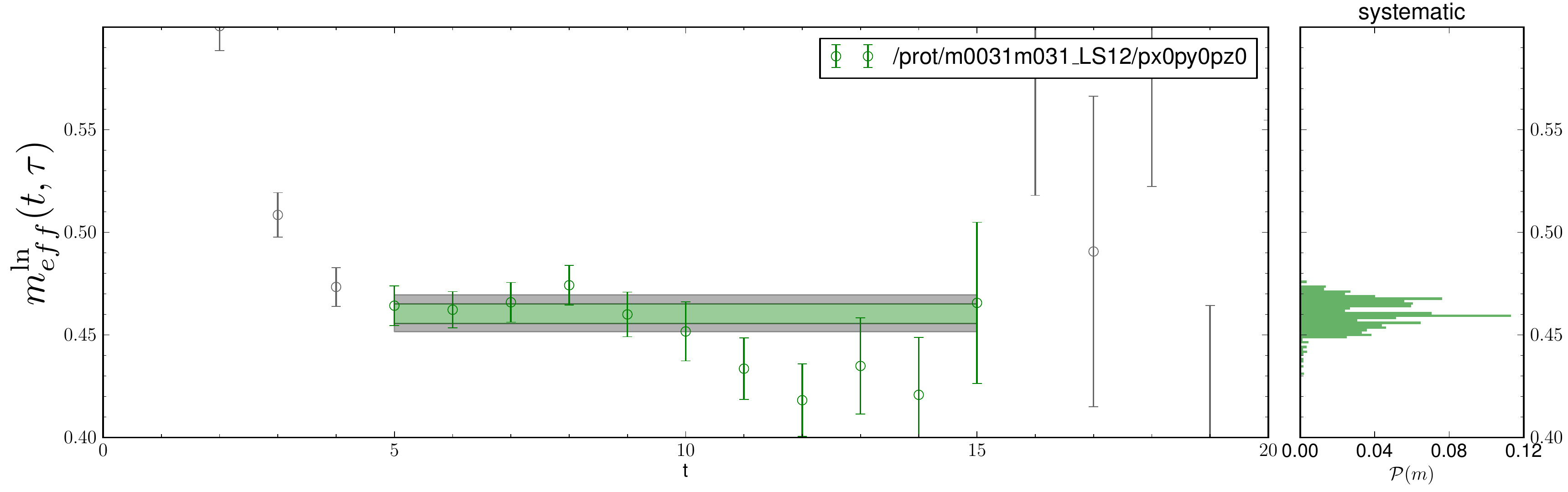}
\caption{\label{fig:mp_fine}Proton mass and representative effective mass plots on the \fine ensembles.}
\end{figure}
%%%%%%%%%%%%%%%%%%%%%%%%%%%%%%%%%%%%%%%%%%%%%%%%%%%
In this work, to determine the fitting systematic, the choices of $t_0$ and $\D t$ are varied over a wide range, with $\D t \gtrsim 0.5$~fm.  For each choice, the ground state principal correlation function is fit with a single exponential, Eq.~\eqref{eq:corr_exp}, over ranges of time $t_i - t_f$, chosen independently of $t_0$ and $\D t$.  
The initial and final times in the fit are also varied over a wide range under the constraint $t_f - t_i \gtrsim 0.5$~fm.
For each fit, the $Q$ value is recorded along with the statistical uncertainty of the fit.  The various fits are then averaged with the weight similar to that of the pions, but also suppressed by the statistical uncertainty of the fit;
\begin{align}
	&\bar{m} = \frac{\sum_i m_i w_i}{\sum_j w_j}&
	&\textrm{with } w_i = \frac{Q_i}{\s_i}\, .&
\end{align}
In this way, the plateaus at later times, with larger uncertainties, and hence larger $Q$ values, do not dominate the determination of the fitting systematic.
The resulting fits are displayed along with effective mass plots of representative Matrix-Prony determinations of the ground state principal correlation function in Figs.~\ref{fig:mp_coarse} and \ref{fig:mp_fine}.
In these figures, the colored effective mass points correspond to the time window over which the Matrix-Prony method is applied in the representative choice of times $t_0$ and $\D t$, while the gray effective mass points lie outside this region.
As is evident, the resulting systematic mass-probability distribution tends not to be Gaussian.  For simplicity, we still take the 16\% and 84\% quantiles to define the systematic uncertainty.  The inner colored bands represent the statistical uncertainty, and the outer gray bands represent the statistical and systematic uncertainties added in quadrature.

%%%%%%%%%%%%%%%%%%%%%%%%%%%%%%%%%%%%%%%%%%%%%%%%%%%
%
%	 scale setting
%
%%%%%%%%%%%%%%%%%%%%%%%%%%%%%%%%%%%%%%%%%%%%%%%%%%%
\subsection{Scale setting}
\noindent
To convert from lattice units to physical units we use the scale setting procedure described in Ref.~\cite{Beane:2011zm}.
The dimensionless lattice results are converted into $r_1$ units with $\frac{r_1}{b}(bm_l,bm_s,\beta)$ determined by the MILC Collaboration on each ensemble.
But importantly, it is not the value computed on a given ensemble that is used; it is rather the values that have been extrapolated to the physical light- and strange-quark mass point, $\frac{r_1}{b}(bm_l^\textrm{phy},bm_s^\textrm{phy},\beta)$, which have also been determined by the MILC Collaboration~\cite{Bazavov:2009bb,Bernard:r1b}, listed here in Table~\ref{tab:mass_results}.
While depending upon reference quark mass values, this amounts to a quark-mass independent scale setting procedure, such that all remaining light- and strange-quark mass dependence of the computed observables is that of interest.
The MILC Collaboration has also determined the physical value of $r_1$,
\begin{equation}\label{eq:r1_phys}
	r_1^\textrm{phy} = 0.31174(20)~\textrm{fm}\, ,
\end{equation}
which is used to then convert all values into physical units, Table~\ref{tab:mass_results_mev}.
%%%%%%%%%%%%%%%%%%%%%%%%%%%%%%%%%%%%%%%%%%
\begin{table}
\caption{\label{tab:mass_results_mev}Computed masses and decay constants converted to MeV with $r_1 = 0.31174(20)$~fm.}
\begin{ruledtabular}
\begin{tabular}{ccccc}
$\beta$& $m_{sea}$& V& $m_\pi$~[MeV]& $m_N$~[MeV] \\
\hline
6.75& m010m030& $20^3\times64\times16$& $380.5(.6)({}^{.9}_{.5})$& $1231(3)({}^{3}_{4})$ \\ 
6.76& m010m050& $20^3\times64\times16$& $386.3(.5)({}^{.8}_{.6})$& -- \\ 
6.76& m010m050& $20^3\times32\times16$& -- & $1267(3)({}^{6}_{5})$ \\
6.79& m030m030& $20^3\times64\times16$& $666.4(.5)(.4)$& $1545(3)({}^{5}_{6})$ \\
6.81& m030m050& $20^3\times32\times16$& $682.7(.5)({}^{.4}_{.2})$& $1591(3)({}^{7}_{6})$\\
%\hline
7.06& m0031m0186& $40^3\times96\times12$& $237.8(0.9)(1.3)$& $1078(15)({}^{26}_{22})$\\
7.08& m0031m031& $40^3\times96\times12$& $241.6(0.8)({}^{2.0}_{1.8})$& $1094(11)({}^{19}_{18})$\\
\end{tabular}
\end{ruledtabular}
\end{table}
%%%%%%%%%%%%%%%%%%%%%%%%%%%%%%%%%%%%%%%%%%%%%%%%%%%

There is an important additional advantage to this method of scale setting.  To invoke the Feynman-Hellmann theorem, the change in the nucleon mass with respect to a change in the strange-quark mass must be undertaken with all other parameters held fixed~\cite{Toussaint:2009pz,Freeman:2012ry}.
The MILC Collaboration chose to make slight changes in the coupling $\beta$ while changing the light quark masses.
Our scale setting procedure allows us to asses the quantitative significance of the slightly different values of $\b$ used on the pairs of ensembles, \{m010m030,m010m050\}, \{m030m030,m030m050\} and \{m0031m0186,m0031m031\}.
For each pair, the relative difference in the values of $\b$ was less than 1\% ($\frac{\b^{(2)}-\b^{(1)}}{\b^{(2)}+\b^{(1)}} < 0.01$) and the corresponding relative difference in the values of  $\frac{r_1}{b}(bm_l^\textrm{phy},bm_s^\textrm{phy},\beta^{(i)})$ are also less than 1\%.
While strictly speaking, the change in $m_s$ was not undertaken with all other parameters held fixed, the effect of this change is contained well within the other uncertainties on the determined values of $m_s \langle N| \bar{s} s | N \rangle$, as detailed in the next section.

%%%%%%%%%%%%%%%%%%%%%%%%%%%%%%%%%%%%%%%%%%%%%%%%%%%
%
%	 ssbar matrix element
%
%%%%%%%%%%%%%%%%%%%%%%%%%%%%%%%%%%%%%%%%%%%%%%%%%%%
\section{The Strange Scalar Matrix Element in the Nucleon\label{sec:sbars}}
\begin{figure}
\includegraphics[width=0.8\textwidth]{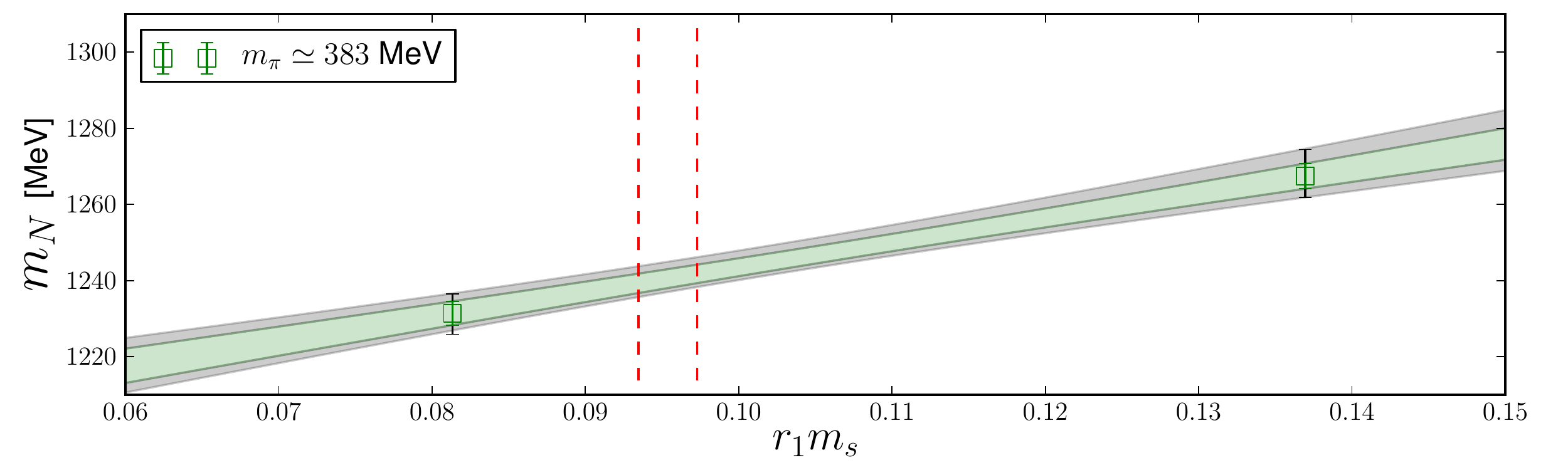}
\includegraphics[width=0.8\textwidth]{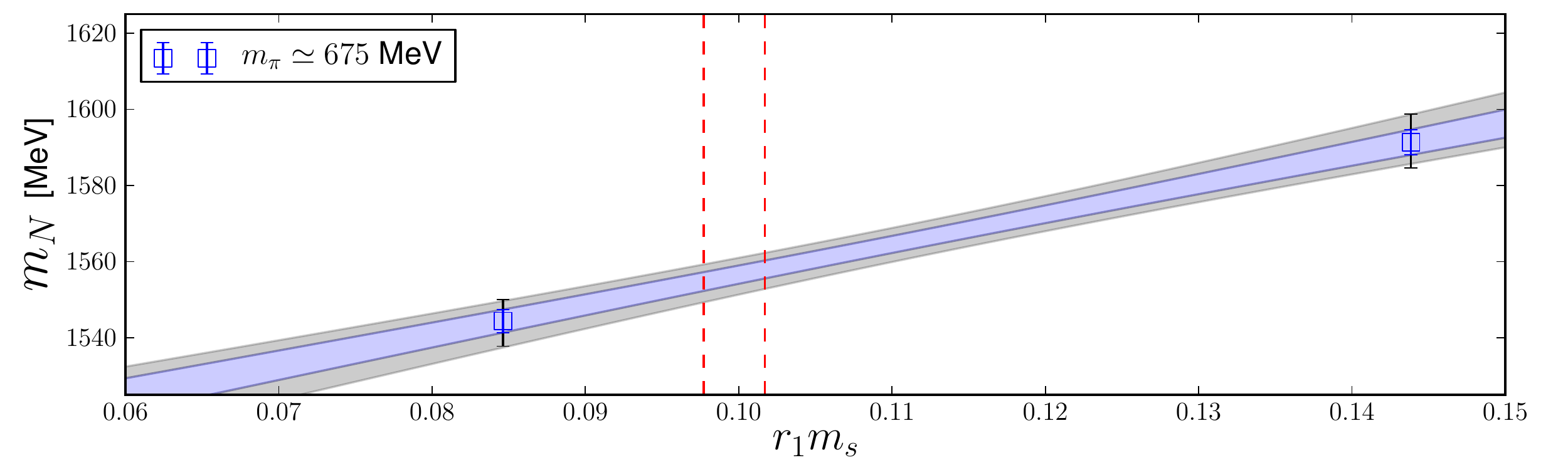}
\includegraphics[width=0.8\textwidth]{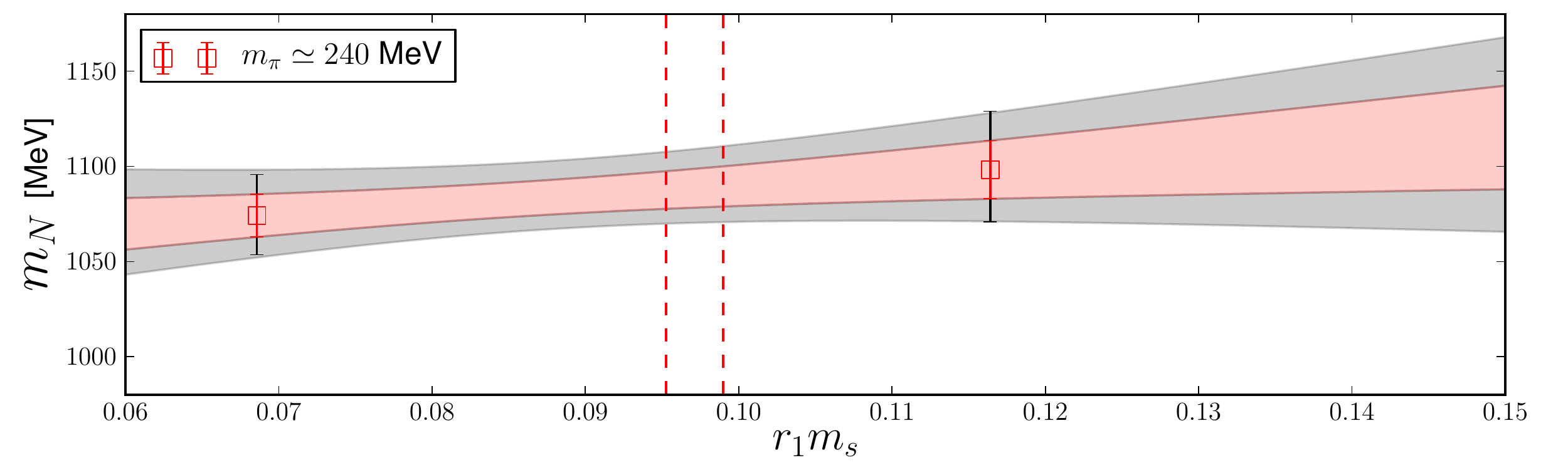}
\caption{\label{fig:mn_vs_ms}Nucleon mass versus the strange-quark mass on the \coarse and \fine ensembles.  The vertical dashed lines represent the 68\% confidence interval for the determination of $bm_s^\textrm{phy}$ on the \coarse and \fine ensembles.  The conversion to $r_1$ units is performed as in Eq.~\eqref{eq:mssbars_mev} using $\frac{1}{2}(\frac{r_1}{b}^{(1)} + \frac{r_1}{b}^{(2)})$ for each pair of ensembles.}
\end{figure}

\noindent
As discussed in the Introduction, there are a few methods for determining the scalar strange-quark matrix element in the nucleon.  These include a direct calculation of the matrix element employed by some groups~\cite{Babich:2010at,Takeda:2010cw,Bali:2011ks,Dinter:2012tt,Gong:2012nw,Oksuzian:2012rzb,Engelhardt:2012gd}, an indirect determination through the Feynman-Hellmann theorem~\cite{Young:2009zb,Durr:2011mp,Horsley:2011wr,Semke:2012gs,Shanahan:2012wh,Ren:2012aj,Jung:2013rz}, Eq.~\eqref{eq:FH}%footnote
\footnote{The first attempt to determine the strange content of the nucleon from lattice QCD with the Feynman-Hellmann method utilized $SU(3)$ baryon $\chi$PT analysis of \coarse MILC results~\cite{Frink:2005ru} resulting in a value consistent with zero.}, 
and a hybrid approach~\cite{Toussaint:2009pz,Freeman:2012ry}.
This work utilizes the Feynman-Hellmann method.
For each light quark mass ensemble, we have a determination of the nucleon mass at values of the strange-quark mass which straddle the physical strange-quark mass.
These results, Table~\ref{tab:mass_results_mev}, can be used to interpolate to the physical value of the strange-quark mass, Taylor expanding about $bm_s^\textrm{phy}$, and determine the two quantities 
\begin{align}
	&m_N(m_s^\textrm{phy})\, ,&
	&\frac{\partial m_N(m_s)}{\partial m_s} \Big|_{m_s^\textrm{phy}}\, .&
\end{align}
%%%%%%%%%%%%%%%%%%%%%%%%%%%%%%%%%%%%%%%%%%
%%  ms < N | sbar s | N >
%%%%%%%%%%%%%%%%%%%%%%%%%%%%%%%%%%%%%%%%%%
\begin{table}
\caption{\label{tab:ms_sbars}Extracted values of $m_N(m_s^\textrm{phy})$ and $m_s^\textrm{phy} \langle N | \bar{s} s | N \rangle$.  The first uncertainty is statistical, the second fitting systematics and the third is from the uncertainty on the determination of $m_s^\textrm{phy}$.}
\begin{ruledtabular}
\begin{tabular}{ccc}
$m_\pi$~[MeV]& $m_N(m_s^\textrm{phy})$~[MeV]& $m_s^\textrm{phy} \langle N | \bar{s} s | N \rangle$~[MeV] \\
\hline
$383.4(.6)({}^{.9}_{.6})$& $1241(2)(3)(1)$& 62(8)(11)(1) \\
$674.6(.5)(.4)$& $1556(2)(4)(2)$& $79(8)(13)(2)$ \\
\hline
$240(1)(2)$& $1090(11)(17)(1)$& $50(40)(65)(1)$
\end{tabular}
\end{ruledtabular}
\end{table}
%%%%%%%%%%%%%%%%%%%%%%%%%%%%%%%%%%%%%%%%%%
To apply the Feynman-Hellmann theorem with all parameters except $m_s$ held (approximately) fixed, the following approximation for the derivative is used,
\begin{equation}\label{eq:mssbars_mev}
	m_s \langle N| \bar{s} s | N \rangle [\textrm{MeV}] =
	\frac{ \frac{r_1}{b}^{(2)} bm_N^{(2)} - \frac{r_1}{b}^{(1)} bm_N^{(1)}}
		{\frac{r_1}{b}^{(2)} bm_s^{(2)} - \frac{r_1}{b}^{(1)} bm_s^{(1)}}
	\times
	\frac{\frac{r_1}{b}^{(2)} + \frac{r_1}{b}^{(1)}}{2}
	b m_s^\textrm{phy}
	\times
	\frac{197.3 \textrm{ MeV fm}}{r_1^\textrm{phy}[\textrm{fm}]}\, ,
\end{equation}
where $\frac{r_1^{(i)}}{b}$ denotes the value of $\frac{r_1}{b}(bm_l^\textrm{phy},bm_s^\textrm{phy},\beta^{(i)})$ for the given ensemble with all parameters except $bm_s$ held approximately fixed and $r_1^\textrm{phy}[\textrm{fm}]$ is taken from Eq.~\eqref{eq:r1_phys}.
The MILC Collaboration has determined values of the strange-quark mass to be $bm_s^\textrm{phy} = 0.0350(7)$ and $bm_s^\textrm{phy} = 0.0261(5)$ on the \coarse and \fine ensembles respectively~\cite{Aubin:2004ck,Bazavov:2009bb}.
The resulting values of $m_N(m_s^\textrm{phy})$ and $m_s^\textrm{phy} \langle N | \bar{s} s | N \rangle$ are collected in Table~\ref{tab:ms_sbars} and the resulting interpolations are displayed in Fig.~\ref{fig:mn_vs_ms}.
%%%%%%%%%%%%%%%%%%%%%%%%%%%%%%%%%%%%%%%%%%
%%%%%%%%%%%%%%%%%%%%%%%%%%%%%%%%%%%%%%%%%%
In these figures, the vertical dashed lines represent the 68\% confidence interval for the determination of $bm_s^\textrm{phy}$ on the \coarse and \fine ensembles.
The uncertainty on $bm_s^\textrm{phy}$ is included in the analysis and represented by the third uncertainty in Table~\ref{tab:ms_sbars}.
The conversion to $r_1$ units is performed as in Eq.~\eqref{eq:mssbars_mev} using $\frac{1}{2}(\frac{r_1}{b}^{(1)} + \frac{r_1}{b}^{(2)})$ for each pair of ensembles.
The estimated correction due to the difference in $\b$ on the pairs of ensembles is at the same level as the uncertainty arising from the determination of $bm_s^\textrm{phy}$, which are at least an order of magnitude smaller than the statistical or other systematic uncertainties.
On the \coarse ensembles, a precise determination of the scalar matrix element is obtained.  However, on the \fine ensembles, the results are too imprecise to determine a nonzero value.

%%%%%%%%%%%%%%%%%%%%%%%%%%%%%%%%%%%%%%%%%%%%%%%%%%%
%
%	 CHIRAL EXTRAPOLATION
%
%%%%%%%%%%%%%%%%%%%%%%%%%%%%%%%%%%%%%%%%%%%%%%%%%%%
\subsection{Chiral extrapolation}
\begin{figure}
\includegraphics[width=0.8\textwidth]{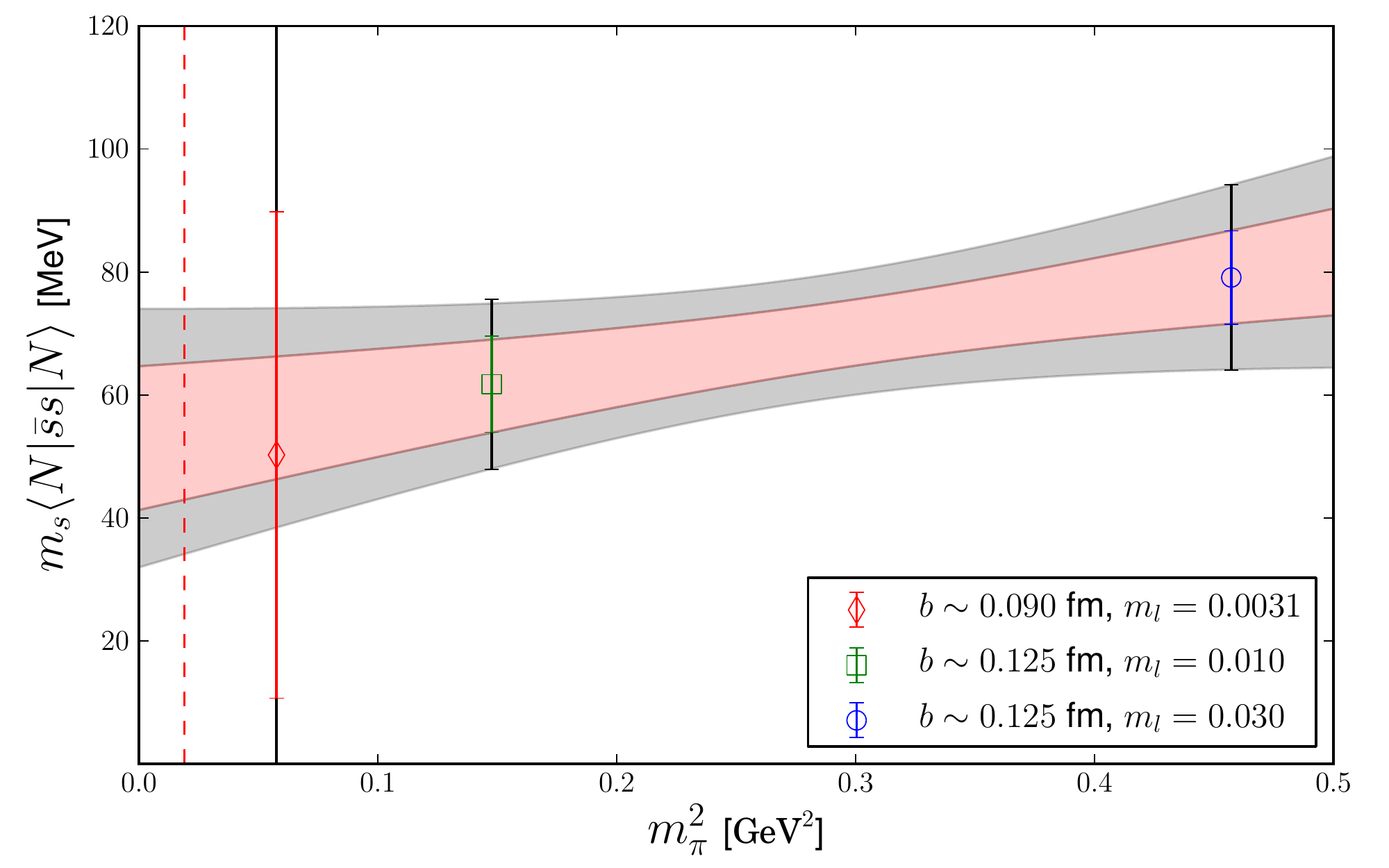}
\caption{\label{fig:ms_sbars_v_mpisq}Light quark extrapolation of $ms\langle N| \bar{s} s | N \rangle$ versus $m_\pi^2$.  The location of the vertical dashed line is given by $(m_\pi^\textrm{phy})^2$.}
\end{figure}

\noindent
The results for $m_s <N|\bar{s}s|N>$ must be extrapolated to the physical value of the pion mass.
In Ref.~\cite{Chen:2002bz}, the two-flavor extrapolation formula for this matrix element was determined at next-to-leading order (NLO) in the chiral expansion,
\begin{equation}
	\langle N | \bar{s} s | N \rangle = 
	<N|\bar{s}s|N>^0
	-\frac{g_{\pi N\D}^2}{4\pi^2 f^2} \left( <N|\bar{s}s|N>^0 - <\D|\bar{s}s|\D>^0 \right)
		\mc{J}_{m_\pi}^\D
	+\tilde{E}_s \frac{m_\pi^2}{8\pi^2 f^2}\, ,
\end{equation}
where $\langle H | \bar{s} s | H \rangle^0$ represent the leading-order (LO) contribution to the scalar strange matrix element in the hadron $H$, $g_{\pi N\D}$ is the axial pion-nucleon-delta coupling appearing in the $SU(2)$ baryon chiral Lagrangian, $\mc{J}_{m_\pi}^\D$ is a chiral loop function nonanalytic in the pion mass and the delta-nucleon mass splitting ($\D=m_\D - m_N$) and $\tilde{E}_s$ is a low-energy constant appearing at NLO.
In the large-$N_c$ expansion, the LO matrix elements for the nucleon and the delta are both $\mc{O}(N_c^{-1})$, but there is no cancellation at this order~\cite{Jenkins:1995gc}, so one does not expect a strong cancellation between these NLO contributions.%footnote
\footnote{See also Ref.~\cite{Cherman:2012eg} for further discussion on the baryon masses in the large $N_c$ counting.} 
In principle, one should use the partially quenched formula, also provided in Ref.~\cite{Chen:2002bz}, and convert it to the relevant mixed-action formula~\cite{Chen:2007ug} to perform the extrapolation.  
However, clearly the most significant shortcoming of the present work is the limited number of light quark mass points.  
With nonzero results at only a single lattice spacing, the mixed-action extrapolation cannot be performed regardless.
The best that can be done with the present results is a simple, effectively zero degree of freedom extrapolation using the formula,
\begin{equation}\label{eq:sbars_simple}
	m_s \langle N | \bar{s} s | N \rangle = c_0 + c_2 m_\pi^2\, .
\end{equation}
While this will not result in a precise and accurate determination of the scalar strange matrix element, it will provide a good guide to the approximate value at the physical point.
While not a rigorous expectation, it has been found that matrix elements of the nucleon tend to have very mild pion mass dependence; see for example the recent review~\cite{Lin:2012ev}.
Performing this simplistic pion mass extrapolation, using the isospin averaged $m_\pi^\textrm{phy} = 138.0$~MeV, we obtain
\begin{equation}
	m_s^\textrm{phy} \langle N | \bar{s} s | N \rangle \Big|_{m_\pi^\textrm{phy}} = 
	54 \pm 11 \pm 17 \textrm{ MeV}\, .
\end{equation}
The extrapolation is displayed in Fig.~\ref{fig:ms_sbars_v_mpisq}.
%%%%%%%%%%%%%%%%%%%%%%%%%%%%%%%%%%%%%%%%%%%%%%%%%%%
%%%%%%%%%%%%%%%%%%%%%%%%%%%%%%%%%%%%%%%%%%%%%%%%%%%

\begin{figure}
\includegraphics[width=0.48\textwidth]{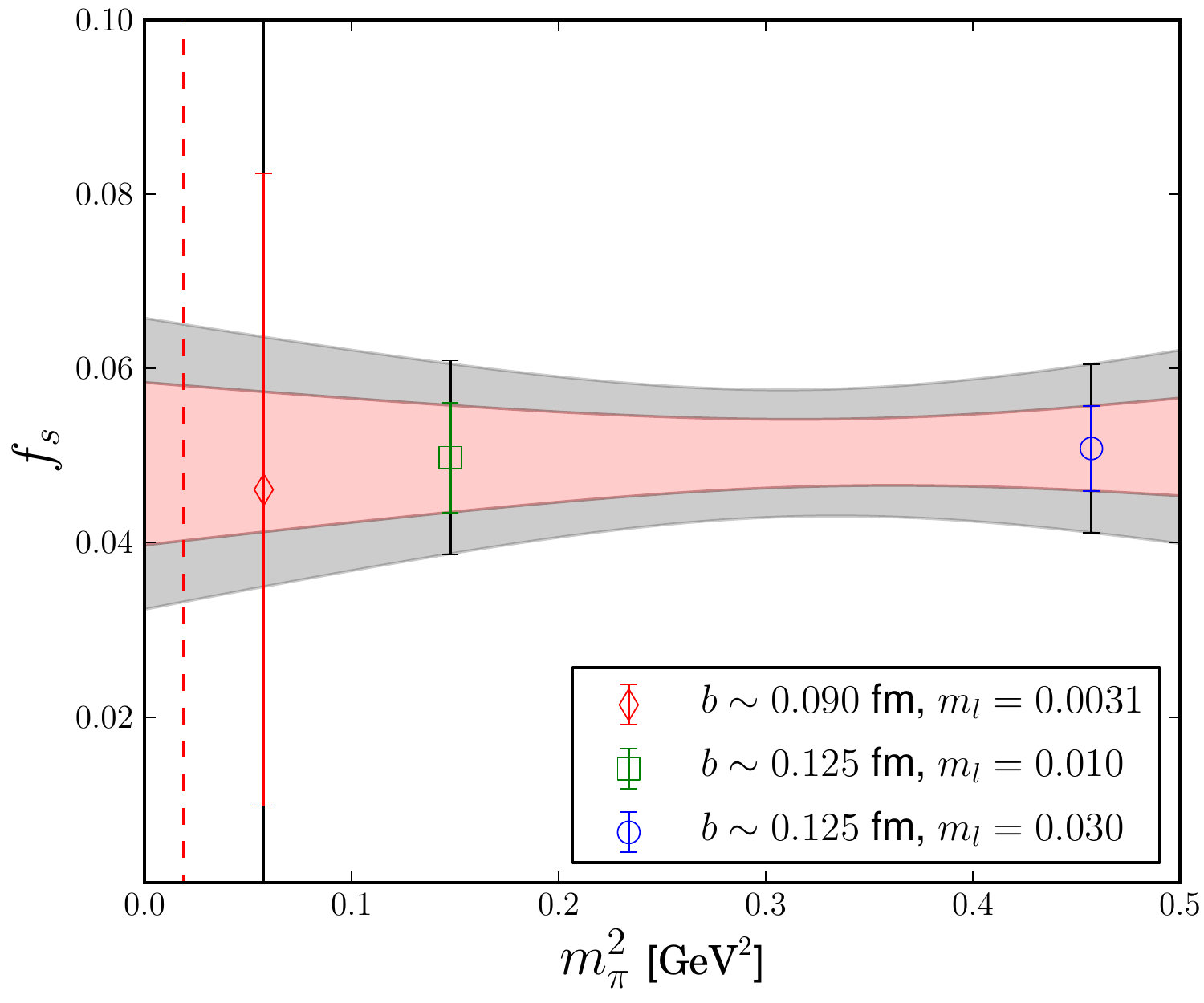}
\includegraphics[width=0.48\textwidth]{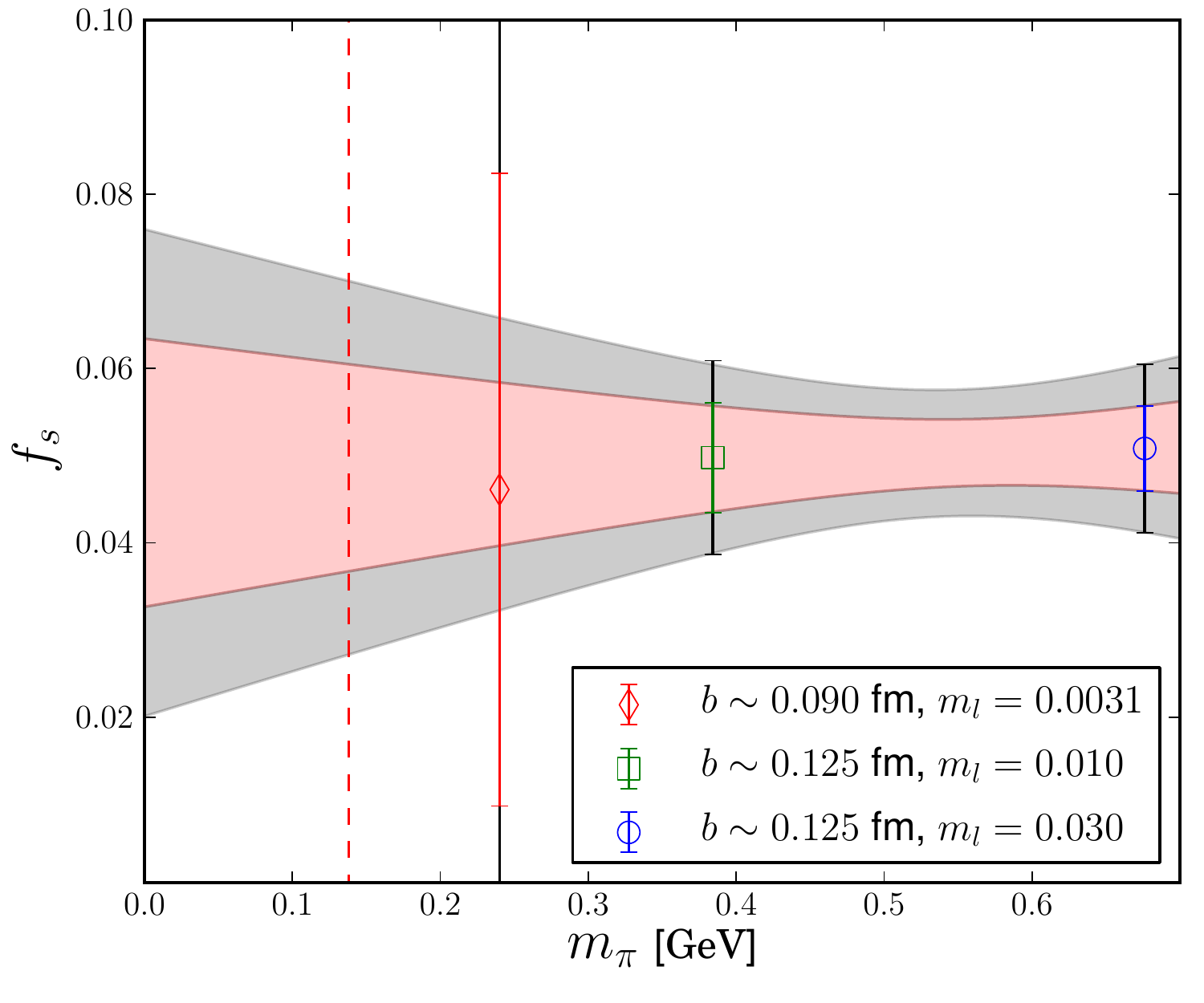}
\caption{\label{fig:fs_extrap}Extrapolation of $f_s$.  The location of the vertical dashed line in each plot is determined from $m_\pi^\textrm{phy}$.}
\end{figure}

Given the limited ability to perform the chiral extrapolation, we also explore the light quark mass dependence of $f_s = m_s \langle N | \bar{s} s | N \rangle / m_N$ to improve the estimate of systematic uncertainties.
It has been observed that the nucleon mass displays a remarkably linear dependence on the pion mass~\cite{WalkerLoud:2008bp,WalkerLoud:2008pj}.
For this reason, the following two extrapolation functions are used to estimate extrapolation systematics:
\begin{subequations}
\begin{align}
f_s &= f_s^{(0)} + f_s^{(2)} m_\pi^2\, ,\label{eq:f_s_mpisq}\\
f_s &= f_s^{(0)} + f_s^{(1)} m_\pi\, ,\label{eq:f_s_mpi}
\end{align}
\end{subequations}
yielding the results
\begin{subequations}
\begin{align}
f_s &= 0.049 \pm 0.009 \pm 0.013\, ,\\
f_s &= 0.049 \pm 0.012 \pm 0.018\, ,
\end{align}
\end{subequations}
respectively.
These extrapolations are displayed in Fig.~\ref{fig:fs_extrap}.  The quantity $f_s$ is observed to have  negligible light quark mass dependence.
%%%%%%%%%%%%%%%%%%%%%%%%%%%%%%%%%%%%%%%%%%%%%%%%%%%
%%%%%%%%%%%%%%%%%%%%%%%%%%%%%%%%%%%%%%%%%%%%%%%%%%%

These results can be compared with the extrapolation of $m_s \langle N | \bar{s} s | N \rangle$ by converting with the isospin averaged nucleon mass $m_N^\textrm{phy} = 938.9$~MeV.
In Table~\ref{tab:final_extrap}, these three different extrapolation results are collected.
Additionally, a correlated weighted average is performed.  To perform the correlated average, Gaussian distributions of the results in Table~\ref{tab:ms_sbars} are created independently for each light quark mass point, with $N_{Gauss} = 10^4$ in all cases.  For each sample, all three extrapolations are performed, preserving the correlations between the fits, with inverse weights given by the statistical and systematic uncertainties on the individual mass points.
For each sample, these three results are then averaged with weights given by the inverse uncertainties from the individual analyses (quoted in Table~\ref{tab:final_extrap}).
This yields the final result
\begin{subequations}
\begin{align}
m_s \langle N | \bar{s} s | N \rangle & = 48 \pm 10 \pm 15 \textrm{ MeV}\, ,\\
f_s &= 0.051 \pm 0.011 \pm 0.016\, .
\end{align}
\end{subequations}
%%%%%%%%%%%%%%%%%%%%%%%%%%%%%%%%%%%%%%%%%%
\begin{table}
\caption{\label{tab:final_extrap}Extrapolated values of $m_s \langle N | \bar{s} s | N \rangle$ and $f_s$.  These results are averaged in a weighted and correlated fashion described in the text.}
\begin{ruledtabular}
\begin{tabular}{cccc}
Quantity Extrapolated& Extrapolation Function& $m_s^\textrm{phy} \langle N | \bar{s} s | N \rangle$~[MeV]& $f_s$\\
\hline
$m_s^\textrm{phy} \langle N | \bar{s} s | N \rangle$ & Eq.~\eqref{eq:sbars_simple}& 
	$56 \pm 12 \pm 17$& $0.059 \pm 0.012 \pm 0.019$ \\
$f_s$& Eq.~\eqref{eq:f_s_mpisq} & $47 \pm\phantom{0}9\pm 13$& $0.050 \pm 0.009 \pm 0.014$ \\
$f_s$& Eq.~\eqref{eq:f_s_mpi} & $47 \pm 12 \pm 17$& $0.050 \pm 0.012 \pm 0.018$ \\
\hline
Correlated Average & -- & $49\pm10\pm15$& $0.053\pm0.011\pm0.016$ 
\end{tabular}
\end{ruledtabular}
\end{table}
%%%%%%%%%%%%%%%%%%%%%%%%%%%%%%%%%%%%%%%%%%

%%%%%%%%%%%%%%%%%%%%%%%%%%%%%%%%%%%%%%%%%%%%%%%%%%%
%
%	 RESULTS AND DISCUSSION
%
%%%%%%%%%%%%%%%%%%%%%%%%%%%%%%%%%%%%%%%%%%%%%%%%%%%
\section{Results and Discussion\label{sec:results}}
\noindent
For the present work, the Feynman-Hellmann theorem was invoked to determine the strange content of the nucleon through a change $m_N$ as the strange-quark mass is varied
\begin{equation*}
 m_s \langle N | \bar{s} s | N \rangle = m_s \frac{\partial m_N}{\partial m_s} \, .
\end{equation*}
By taking care to set the scale using values of $r_1/b$, which were extrapolated to the physical values of the light- and strange-quark masses, the nucleon mass variation was determined with all other parameters held constant (with precision better than 1\%), as is required for a proper determination of this quantity~\cite{Toussaint:2009pz,Freeman:2012ry}.
There are several groups who have used the Feynman-Hellmann theorem~\cite{Young:2009zb,Durr:2011mp,Horsley:2011wr,Oksuzian:2012rzb,Semke:2012gs,Shanahan:2012wh,Ren:2012aj,Jung:2013rz}
as well as more determinations with a direct calculation of the matrix element~\cite{Babich:2010at,Takeda:2010cw,Bali:2011ks,Dinter:2012tt,Gong:2012nw,Oksuzian:2012rzb,Engelhardt:2012gd} and results from a hybrid approach~\cite{Toussaint:2009pz,Freeman:2012ry}.
Before making a detailed comparison with other works, we first highlight advantages and disadvantages of the present work.
The distinct advantage of using the Feynman-Hellmann theorem over direct methods is that the ground state plateau of the nucleon can be significantly more reliably determined than the plateau for the matrix element calculation with equal computing resources; see the plots of ratio determinations in any of Refs.~\cite{Babich:2010at,Takeda:2010cw,Bali:2011ks,Dinter:2012tt,Gong:2012nw,Oksuzian:2012rzb,Engelhardt:2012gd} (the direct calculation requires a vacuum subtraction, adding substantial statistical noise).
The disadvantage of most groups employing the Feynman-Hellmann theorem is the reliance upon $SU(3)$ baryon $\chi$PT~\cite{Young:2009zb,Semke:2012gs,Shanahan:2012wh,Ren:2012aj}, which is known to not have a converging expansion for the nucleon mass~\cite{WalkerLoud:2008bp,Jenkins:2009wv,Torok:2009dg,Ishikawa:2009vc,WalkerLoud:2011ab}.  
Therefore, it is not clear that the full extrapolation systematic has been properly addressed in those works.%footnote
\footnote{The work in Ref.~\cite{Durr:2011mp} also uses $SU(3)$ baryon $\chi$PT, but uses a variety of other extrapolation methods, resulting in a conservative estimate of their uncertainties.} 
This concern is substantiated by the discrepancy between independent $SU(3)$ baryon $\chi$PT analyses and their determination of $f_s$~\cite{Young:2009zb,Shanahan:2012wh,Semke:2012gs,Ren:2012aj}.%footnote
\footnote{Despite these criticisms, we point out in Ref.~\cite{Shanahan:2012wh}, a striking agreement is found between baryon mass results extrapolated from one set of lattice calculations~\cite{WalkerLoud:2008bp,Ishikawa:2009vc}, with $SU(3)$ baryon $\chi$PT, and then used to predict results from a completely independent calculation~\cite{Bietenholz:2011qq}.  Moreover, independent verification of the consistency of various lattice calculations of the ground state baryon spectrum and $SU(3)$ baryon $\chi$PT has been found~\cite{Semke:2011ez,Semke:2012gs,Lutz:2012mq}.
} 
For further discussion on the convergence problems using $SU(3)$ baryon $\chi$PT specifically for the scalar strange content of the nucleon, see Ref.~\cite{Alarcon:2012nr}.
The current work does not suffer from this issue.

The most severe limitation of the present work is the small number of light quark mass points (two) for which there is a nonzero determination of $m_s \langle N | \bar{s} s | N \rangle$.  Given the significant numerical cost of the domain-wall propagators on the \fine ensemble with $m_\pi \simeq 240$~MeV, it is not clear how soon a more precise determination will be obtained at this point.
Given the very mild light quark mass dependence observed in this work, and in nucleon matrix elements in general, we believe the present determination offers a reliable estimate of the scalar strange content of the nucleon, but neither a precise nor demonstrably accurate value.  Our final result is
\begin{align*}
m_s \langle N | \bar{s} s | N \rangle & = 49 \pm 10 \pm 15 \textrm{ MeV}\, ,\\
f_s &= 0.053 \pm 0.011 \pm 0.016\, .
\end{align*}

%%%%%%%%%%%%%%%%%%%%%%%%%%%%%%%%%%%%%%%%%%%%%%%%%%%
%
%	 LATTICE AVERAGE
%
%%%%%%%%%%%%%%%%%%%%%%%%%%%%%%%%%%%%%%%%%%%%%%%%%%%
\subsection{Lattice QCD comparison and average}
\noindent
Given the phenomenological importance of the scalar strange content of the nucleon, see for example Refs.~\cite{Bottino:1999ei,Bottino:2001dj,Kaplan:2000hh,Ellis:2008hf,Ellis:2009ai,Giedt:2009mr,Freytsis:2010ne,Hill:2011be,Cheung:2012qy}, it is prudent to review the limitations of the present determination and to compare and contrast these results to other lattice QCD determinations.
There are two results which use the same MILC ensembles with staggered valence quarks~\cite{Toussaint:2009pz,Freeman:2012ry} and one determination with the same mixed-action scheme but a direct determination~\cite{Engelhardt:2012gd}.
It is interesting to first compare our results with these.  
Reference~\cite{Freeman:2012ry} (an update of \cite{Toussaint:2009pz}) quotes only the value of $\langle N | \bar{s} s | N \rangle$ in $\overline{\textrm{MS}}\ (2 \textrm{ GeV})$.
To convert this number into the dimensionful, renormalization scheme invariant quantity, we take the ratio of quoted values $m_s \langle N | \bar{s} s | N \rangle / \langle N | \bar{s} s | N \rangle$ from Ref.~\cite{Toussaint:2009pz}, which amounts to $m_s[\overline{\textrm{MS}}\ (2 \textrm{ GeV})] = 86$~MeV.
Alternatively, we could use the strange-quark mass determination of HPQCD~\cite{Mason:2005bj} (updated by MILC~\cite{Bazavov:2009tw}), $m_s[\overline{\textrm{MS}}\ (2 \textrm{ GeV})] = 89.0(4.8)$~MeV, but within uncertainties, these are the same.
Comparing to these works, as well as the mixed-action calculation, good agreement is found:
\begin{equation}
m_s \langle N | \bar{s} s | N \rangle [\textrm{MeV}]=
	\left\{ \begin{array}{ll}
		59\pm6\pm8 & \textrm{Ref.~\cite{Toussaint:2009pz}}\\
		54\pm5\pm6 & \textrm{Ref.~\cite{Freeman:2012ry}}\\
		43\pm8\pm6 & \textrm{Ref.~\cite{Engelhardt:2012gd}}\\
		49\pm10\pm15 & \textrm{ present work}
		\end{array}
		\right. \, .
\end{equation}

In the literature, there is currently no determination of $f_s$ that considers all the available results from lattice QCD, and so we take the opportunity to provide one here.%footnote
\footnote{There is a recent review on the topic in Ref.~\cite{Young:2013nn}, but a lattice average is not provided.} 
We use an approach similar to the FLAG working group of FLAVIANET, which has provided lattice determinations of various quantities important to low-energy hadronic physics~\cite{Colangelo:2010et}.
In particular, the FLAG working group has developed a scheme to judge the confidence to place in various determinations, based upon standards such as the lightest pion mass used, whether or not a continuum limit has been performed, and whether the infinite volume limit has been performed.  
For each criterion, a green star ({\large{\color{green}{$\star$}}}) is awarded to results that meet the strictest constraints, an orange circle ({\large{\color{orange}{$\bullet$}}}) is given to results with room for improvement and a red square ({\scriptsize{\color{red}{$\blacksquare$}}}) to those with room for significant improvement.
This provides a useful guide to people outside the lattice community and motivation for those in the community to improve their results.

\begin{figure}
\includegraphics[width=0.7\textwidth]{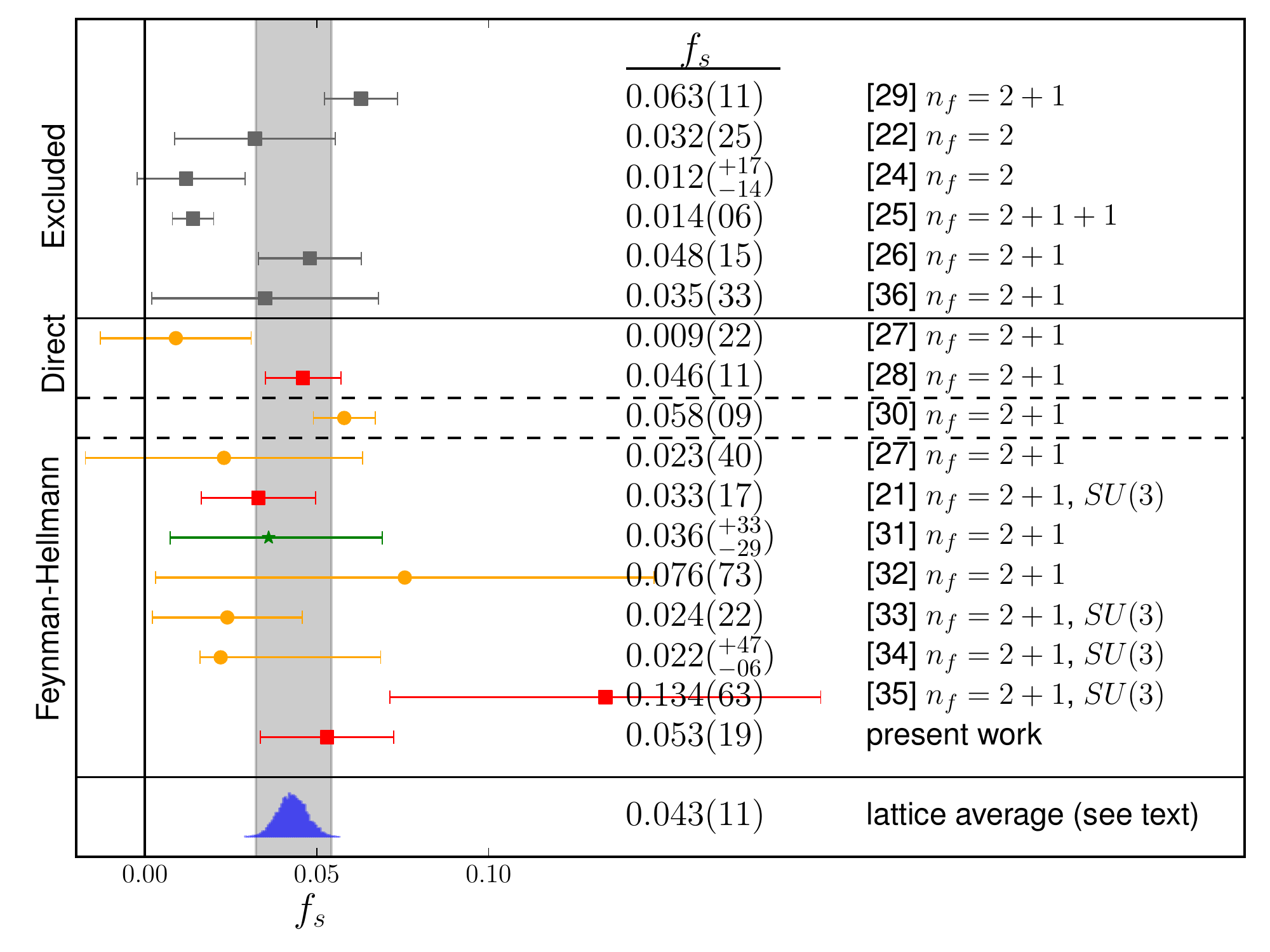}
\caption{\label{fig:fs_compare}Comparison and average of lattice QCD calculations of $f_s$ as described in the text.  
Only values that have been extrapolated to the physical quark masses are used.  Results that quote $m_s \langle N | \bar{s} s | N \rangle$ are normalized by $m_N = 938.9$~MeV to convert to $f_s$.
The quoted uncertainties are taken as the statistical and systematic uncertainties added in quadrature from a given reference.  $n_f=2+1$ indicates a dynamical strange quark as well as up and down.  $SU(3)$ is used to indicate results that rely heavily on $SU(3)$ baryon $\chi$PT.
Some results are excluded for various reasons but displayed to demonstrate their consistency:
\cite{Toussaint:2009pz} was updated in \cite{Freeman:2012ry},
the $n_f=2$ results~\cite{Takeda:2010cw,Bali:2011ks} were not averaged with the $n_f=2+1$,
the results in \cite{Dinter:2012tt} were preliminary and not extrapolated to the physical pion mass,
the results in \cite{Gong:2012nw,Jung:2013rz} are preliminary and only exist in a conference proceedings.
All excluded results are presented as quoted in the literature, with no attempt to perform chiral extrapolations
}
\end{figure}

Using the standards of Ref.~\cite{Colangelo:2010et}, most results for $f_s$ receive an orange circle.  There is one group that receives the green star, and the rest receive a red square.  
The results with a red square suffer either from too few light quark mass points to make a reliable chiral extrapolation or they rely too heavily on $SU(3)$ baryon $\chi$PT.
There are two analyses that we promote from a red square to an orange circle because while they rely heavily on $SU(3)$ baryon $\chi$PT, they have demonstrated a remarkable consistency of their analysis with four or more independent lattice calculations~\cite{Semke:2012gs,Shanahan:2012wh}.
We exclude results that are either not published or not in an arXiv e-print posting (as results in conference proceedings often undergo larger-than-quoted systematic changes).
We further exclude results which have not been extrapolated to the physical value of the light-quark mass, and results calculated without dynamical strange quarks ($n_f=2$) are not included in the average.
To convert results from $m_s \langle N| \bar{s} s | N \rangle$ to $f_s$, we use $m_N = 938.9$~MeV.
These results are displayed in Fig.~\ref{fig:fs_compare}.

For the scalar strange content of the nucleon, the current state of results is such that a simple weighted average of good (green star) results can not be performed in a meaningful way.
As can be seen in Fig.~\ref{fig:fs_compare}, there is good consistency between most of the results.  
There are not a large number of orange circle results, so we chose to include all results in the average.
Moreover, we believe despite their red-square assignment, these results offer valuable information which should not be ignored at this time.

%%%%%%%%%%%%%%%%%%%%%%%%%%%%%%%%%%%%%%%%%%%%%%%%%%%
%%%%%%%%%%%%%%%%%%%%%%%%%%%%%%%%%%%%%%%%%%%%%%%%%%%
A simple weighted average, using the quoted uncertainties as the inverse weights, produces an unbelievably small final uncertainty.
This also ignores the fact that systematic uncertainties are typically non-Gaussian, and in the case of lattice QCD calculations, not cleanly separable from the statistical uncertainties.
Moreover, it does not account for the quality of the results, judged using the rubric of the FLAG working group.
In an attempt to include all these issues, the following \textit{ad hoc} procedure is used to perform a weighted average of all the results (presented in Figure~\ref{fig:fs_compare}):
\begin{enumerate}[i)]
\item for each of the $N_{latt}=11$ results, $f_i \pm \s_i^\pm$, an independent random sample is generated with a sample size of $N_{dist}=10^4$, drawn from a uniform distribution between the quoted uncertainties,
\begin{align*}
&\textrm{for i in range($N_{latt}$):}\\
&\quad\textrm{for j in range($N_{dist}$):}\\
&\quad\quad f_{i,j} = \textrm{random.uniform}(f_i - \s_i^-, f_i + \s_i^+)
\end{align*}

\item for each random sample, a weighted average of all results is performed, with weight 
\begin{equation}\label{eq:avg_weights}
	w_i = y_i / \s_i\, ,
\end{equation}
where $\s_i$ is the symmetric uncertainty, $\s_i = 0.5*(\s_i^+ + \s_i^-)$ from a given result, and we arbitrarily chose $y_i = 1,2,3$ for the red square, orange circle and green star, respectively.  An extra multiplicative reduction of $0.5$ is assigned to results which rely heavily on $SU(3)$ baryon $\chi$PT,
\begin{align*}
&\textrm{for j in range($N_{dist}$):}\\
&\quad \bar{f}_j = \frac{\sum_i w_i\ f_{i,j}}{\sum_{i^\prime} w_{i^\prime}}
\end{align*}
The choice to weight with $1/\s_i$ instead of $1/\s_i^2$ is partly motivated from the non-Gaussian behavior of the systematic uncertainties that typically dominate the lattice results.

\item the mean and 99\% confidence intervals of the resulting distribution are quoted, see Fig.~\ref{fig:fs_compare}

\end{enumerate}

A principal concern one should have about this average is the choice of weights used, Eq.~\eqref{eq:avg_weights}.
To help judge the stability of the average presented here, a variety of different weights are chosen, and the subsequent averages are compared and presented in Table~\ref{tab:fs_weights}.  
The different choices in weights result in very consistent values.
This is a statement about the consistency of the values of $f_s$ from a variety of lattice QCD calculations, and it is this striking consistency that leads us to believe a lattice average with the present results is meaningful (despite the shortcomings of most of the individual results).
The resulting lattice average, quoted at the 99\% confidence interval to be conservative, is
\begin{align}\label{eq:fs_avg}
m_s \langle N | \bar{s} s | N \rangle &= 40 \pm 10 \textrm{ MeV}\, ,\nonumber\\
f_s &= 0.043 \pm 0.011\, .
\end{align}

%%%%%%%%%%%%%%%%%%%%%%%%%%%%%%%%%%%%%%%%%%
\begin{table}
\caption{\label{tab:fs_weights} Value of $f_s$ determined with various weights as described in text.  The right-most value (with $w_i = y_i/\s_i$) is the value taken in this work to represent the lattice average.}
\begin{ruledtabular}
\begin{tabular}{c|ccccc|c}
$w_i$ &$1/\s_i^2$& $y_i/\s_i^2$& $1/\s_i$& $y_i^2/\s_i$& 1& $y_i/\s_i$ \\
\hline
$f_s(68\%)$ &0.0458(31)& 0.0470(35)& 0.0442(36)& 0.0420(55)& 0.0487(63)& 0.0428(41)
\end{tabular}
\end{ruledtabular}
\end{table}
%%%%%%%%%%%%%%%%%%%%%%%%%%%%%%%%%%%%%%%%%%

As was first discussed in Refs.~\cite{Young:2009zb,Giedt:2009mr}, there is now compelling evidence from lattice QCD that the value of the scalar strange content of the nucleon is substantially smaller than previously estimated and does not play as significant a role in dark-matter searches as previously thought~\cite{Bottino:1999ei,Bottino:2001dj,Ellis:2008hf,Hill:2011be}.
This has potential implications for the importance of spin-dependent dark-matter searches as discussed in Ref.~\cite{Freytsis:2010ne}.
For a recent review of the lattice QCD determinations of the scalar strange content of the nucleon, see Ref.~\cite{Young:2013nn}.

%%%%%%%%%%%%%%%%%%%%%%%%%%%%%%%%%%%%%%%%%%%%%%%%%%%
%
%	 c,b,t matrix elements
%
%%%%%%%%%%%%%%%%%%%%%%%%%%%%%%%%%%%%%%%%%%%%%%%%%%%
\subsection{Estimating the heavy quark matrix elements}
\noindent
Knowledge of $f_u$, $f_d$ and $f_s$ can be used to determine the values of $f_c$, $f_b$ and $f_t$~\cite{Shifman:1978zn,Kryjevski:2003mh}.
In Ref.~\cite{Kryjevski:2003mh}, these heavy quark matrix elements were computed using perturbative QCD to $\mc{O}(\a_s^3)$, finding%
%FOOTNOTE
\footnote{We have updated the values of the quark masses used in Ref.~\cite{Kryjevski:2003mh} to the current PDG values~\cite{Beringer:1900zz}.}
\begin{align}
&f_c = 0.08896(1-x_{uds})\, ,&
&f_b = 0.08578(1-x_{uds})\, ,&
&f_t = 0.08964(1-x_{uds})\, ,&
\end{align}
where
\begin{equation}
x_{uds} = f_u + f_d + f_s\, .
\end{equation}
The light-quark matrix elements are given by the pion-nucleon sigma term $m_N(f_u+f_d) = \s_{\pi N}$, which has also been determined from lattice QCD.  As can be seen in Ref.~\cite{Young:2013nn}, the determination by the BMW Collaboration~\cite{Durr:2011mp} not only would have the only green-star ranking but also is a good approximation for the average of all lattice QCD calculations of this quantity, with a value $\s_{\pi N} = 39({}^{+18}_{-8})$~MeV.  Combining this with our estimate for $f_s$ yields a value $x_{uds} = 0.085({}^{+.022}_{-.014})$, and values of the heavy-quark matrix elements
\begin{align}
&f_c = 0.0814({}^{+12}_{-20})\, ,&
&f_b = 0.0785({}^{+12}_{-19})\, ,&
&f_t = 0.0820({}^{+13}_{-20})\, ,&
\end{align}
or in dimensionful units
\begin{align}
m_c \langle N | \bar{c} c | N \rangle &= 76({}^{+11}_{-19}) \textrm{ MeV},
\nonumber\\
m_b \langle N | \bar{b} b | N \rangle &= 74({}^{+11}_{-18}) \textrm{ MeV},
\nonumber\\
m_t \langle N | \bar{t} t | N \rangle &= 77({}^{+12}_{-19}) \textrm{ MeV}.
\end{align}
The resulting charm-quark matrix element is in good agreement with the direct lattice QCD calculations of this quantity~\cite{Freeman:2012ry,Gong:2013vja}.

\acknowledgments

\noindent 
PMJ would like to especially thank S.~Beane for many helpful conversations and for suggesting this project.
PMJ also thanks the hospitality of LBNL where some of this work was completed.
AWL would like to thank R.~Lebed for clarifying some subtleties in the large $N_c$ expansion.  AWL would also like to thank J.~Ruderman for helpful conversations.
We thank our fellow members of the NPLQCD Collaboration for providing some of the numerical results used in the present work and for helpful comments.
We thank C.~Bernard for providing the updated values of $r_1/b$ for the MILC Collaboration.
We thank J.~Ruderman for the motivation to compare all lattice results of this quantity.
Numerical calculations for the present work were performed with the 
\texttt{CHROMA} software suite~\cite{Edwards:2004sx}.
We acknowledge computational support from the USQCD SciDAC project, LLNL, the Argonne Leadership Computing Facility at Argonne National Laboratory (Office of Science of the DOE, under Contract No. DE-AC02-06CH11357).  Calculations were also performed on Endeavour, a UNH computing cluster.
The work of PMJ was supported in part by NSF Grant No PHY1206498.
The work of AWL was supported in part by the Director, Office of Energy Research, Office of High Energy and Nuclear Physics, Divisions of Nuclear Physics, of the U.S. DOE under Contract No. DE-AC02-05CH11231.

\bibliography{SU2_MA}

\end{document}